\documentclass[twocolumn]{aastex63}

\newcommand{\msun}{${\rm M}_\sun$}
\newcommand{\lsun}{${\rm L}_\sun$}

\submitjournal{ApJ}

\shorttitle{Kinematics of G335 MM1 ALMA1}
\shortauthors{Olguin et al.}
\graphicspath{{./}{figures/}}
\defcitealias{2021ApJ...909..199O}{Paper I}
\turnoffedit 

\begin{document}

\title{Digging into the Interior of Hot Cores with ALMA (DIHCA). II.\\Exploring the Inner Binary (Multiple) System Embedded in G335 MM1 ALMA1}

\correspondingauthor{Fernando Olguin}
\email{folguin@phys.nthu.edu.tw}

\author[0000-0002-8250-6827]{Fernando A. Olguin} %
\affil{Institute of Astronomy and Department of Physics, National Tsing Hua University, Hsinchu 30013, Taiwan} %

\author[0000-0002-7125-7685]{Patricio Sanhueza} %
\affiliation{National Astronomical Observatory of Japan, National Institutes of Natural Sciences, 2-21-1 Osawa, Mitaka, Tokyo 181-8588, Japan}
\affil{Department of Astronomical Science, SOKENDAI (The Graduate University for Advanced Studies), 2-21-1 Osawa, Mitaka, Tokyo 181-8588, Japan}

\author[0000-0001-6431-9633]{Adam Ginsburg}
\affil{Department of Astronomy, University of Florida, P.O. Box 112055, Gainesville, FL, USA}

\author[0000-0002-9774-1846]{Huei-Ru Vivien Chen} %
\affil{Institute of Astronomy and Department of Physics, National Tsing Hua University, Hsinchu 30013, Taiwan} %

\author[0000-0003-2384-6589]{Qizhou Zhang}
\affiliation{Center for Astrophysics $|$ Harvard \& Smithsonian, 60 Garden Street, Cambridge, MA 02138, USA}

\author[0000-0003-1275-5251]{Shanghuo Li}
\affiliation{Korea Astronomy and Space Science Institute, 776 Daedeokdae-ro, Yuseong-gu, Daejeon 34055, Republic of Korea}

\author[0000-0003-2619-9305]{Xing Lu}
\affiliation{Shanghai Astronomical Observatory, Chinese Academy of Sciences, 80 Nandan Road, Shanghai 200030, People’s Republic of China}

\author[0000-0003-4521-7492]{Takeshi Sakai}
\affiliation{Graduate School of Informatics and Engineering, The University of Electro-Communications, Chofu, Tokyo 182-8585, Japan.}


\begin{abstract}
We observed the high-mass protostellar core G335.579--0.272 ALMA1 at ${\sim}200$\,au (0\farcs05) resolution with the Atacama Large Millimeter/submillimeter Array (ALMA) at 226\,GHz (with a mass sensitivity of  $5\sigma=0.2$\,\msun\ at 10\,K). 
We discovered that at least a binary system is forming inside this region, with an additional nearby bow-like structure (${\lesssim}1000$\,au) that could add an additional member to the stellar system. 
These three sources are located at the center of the gravitational potential well of the ALMA1 region and the larger MM1 cluster. The emission from CH$_3$OH (and many other tracers) is extended ($>1000$\,au), revealing a common envelope toward the binary system. 
We use CH$_2$CHCN line emission to estimate an inclination angle of the rotation axis of 26\degr\ with respect to the line of sight based on geometric assumptions and derive a kinematic mass of the primary source (protostar+disk) of 3.0\,\msun\ within a radius of 230\,au.
Using SiO emission, we find that the primary source drives the large scale outflow revealed by previous observations.
Precession of the binary system likely produces a change in orientation between the outflow at small scales observed here and large scales observed in previous works. 
The bow structure may have originated by entrainment of matter into the envelope due to widening or precession of the outflow, or, alternatively, an accretion streamer dominated by the gravity of the central sources. An additional third source, forming due to instabilities in the streamer, cannot be ruled out as a temperature gradient is needed to produce the observed absorption spectra.

\end{abstract}


\section{Introduction}
\label{sec:intro}

High-mass stars are born in clusters or associations of stars.
They are thus likely to form binary or multiple stellar systems.
During the gravitational collapse of a molecular cloud the initial physical conditions define how the cloud will fragment.
Core fragmentation can create bound systems which can ultimately result in wide binary systems (e.g., \citealp{2007ApJ...656..959K} with wider fragments forming at distances larger than 1000\,au).
On the other hand, when the cores have evolved enough that a disk forms, gravitational instabilities allow the development of substructures in the disk, e.g., spiral arms. 
These can sporadically feed the embedded protostar  \citep[e.g.,][]{2018MNRAS.473.3615M} or aid the formation of additional companions to the central object if they fragment and grow to become gravitational unstable (e.g., \citealp{2021A&A...652A..69M} with fragments forming within 1000\,au).
Finally, when the stars form a cluster and the system relaxes, close encounters can also allow the formation of binary or multiple systems  \citep[e.g.,][]{2012ApJ...754...71K}.

Given the relatively larger distances of high-mass star-forming regions, resolving core and disk fragmentation requires high angular resolution observations achievable only with the current generation of interferometers, such as the Atacama Large Millimeter/submillimeter Array (ALMA).
Observations of resolved single (e.g., G345.4938+01.4677, \citealp{Guzman2020}; AFGL 4176, \citealp{2020ApJ...896...35J}; G17.64+0.16, \citealp{2019A&A...627L...6M}) and binary systems (e.g., IRAS 07299--1651, \citealp{2019NatAs...3..517Z}; IRAS 16547--4247, \citealp{2020ApJ...900L...2T}; and potentially W33A, \citealp{2017MNRAS.467L.120M}) hosting high-mass protostars have shown a diversity of environments.
The observations of high-mass binary systems show that each component has its individual accretion disk which is in turn fed by a circumbinary disk.
Additional substructures detected towards these sources are large scale streamers at $>1000$\,au scales as revealed by the 1.3\,mm continuum of IRAS 07299--1651 \citep{2019NatAs...3..517Z} and of W33A as revealed by 0.8\,mm continuum and molecular line emission \citep{2018MNRAS.478.2505I}, and outflow cavity walls from the continuum observations of IRAS 16547--4247 \citep{2020ApJ...900L...2T}.

As part of the Digging into the Interior of Hot Cores with ALMA (DIHCA) survey, we are studying the prevalence of binary systems in a sample of 30 high-mass star-forming regions (P. Sanhueza, in prep.).
In our first case study of the survey \citep[][hereafter Paper I]{2021ApJ...909..199O}, we analyzed ALMA 1.33\,mm observations of the high-mass source G335.579--0.272 MM1 \citep[distance $d=3.25$\,kpc][hereafter G335 MM1]{}.
These observations revealed 5 continuum sources with 2 of them associated with radio emission observed by \citet{2015A&A...577A..30A}: ALMA1 (radio MM1a) and ALMA3 (radio MM1b).
The most massive source is ALMA1 with an estimated mass of its gas reservoir of 6.2\,\msun\ \citepalias[][]{2021ApJ...909..199O}.
Its radio emission has a spectral index whose origin can be attributed to a hyper-compact \ion{H}{2} region \citep[HC \ion{H}{2};][]{2015A&A...577A..30A}.
Previous ALMA observations of ALMA1 show that the matter around the central source is infalling and rotating at large scales and expanding within the central region \citepalias[][]{2021ApJ...909..199O}.
This source is driving a molecular outflow with an inclination angle, $i$, between 57\degr\ and 76\degr\ as derived from the outflow geometry \citep{2021A&A...645A.142A}, and a position angle ${\rm P.A.}{\sim}210$\degr\ \citepalias[][]{2021ApJ...909..199O}.
We note that when G335 MM1 was originally discovered \citep{Peretto13}, this object was recognized as one of the most massive cores in the Galaxy contained in a 10,000 au diameter \citep[see also][]{Stephens15}.
Therefore, one of the goals of the DIHCA survey is to reveal what is hidden deeply embedded in massive cores and whether they monolithicly form high-mass stars or fragment in binary (multiple) systems. 

We present here high-resolution ALMA observations that resolve substructures within the high-mass core G335 MM1 ALMA1.
Section~\ref{sec:obs} describes the observations.
The results are presented in Section~\ref{sec:results}.
We discuss the origin of the substructures in Section~\ref{sec:discussion}.
Finally, our conclusions are presented in Section~\ref{sec:conclusions}.

\section{Observations}
\label{sec:obs}

We observed G335 MM1 with ALMA at 226.2\,GHz (1.33\,mm) during July 2019 (Project ID: 2016.1.01036.S; PI: Sanhueza).
The 42 antennas of the 12\,m array covered a baseline range of 92.1 to 8500\,m in a configuration similar to C43-8 (hereafter extended configuration). 
As a result, the resolution of the observations was ${\sim} 0\farcs05$ (${\sim}160$\,au) with a maximum recoverable scale (MRS) of 0\farcs78\footnote{Calculated using equation 7.7 from the ALMA Technical Handbook (\url{https://almascience.nao.ac.jp/documents-and-tools/cycle8/alma-technical-handbook}):
\begin{equation}
    \theta_{\rm MRS} \approx \frac{0.983\lambda}{L_5}
\end{equation}
with $\lambda=1.33$\,mm and $L_5$ the 5th percentile baseline length.
} and primary beam FWHM of 25\farcs1. 
The observations were performed in single-pointing mode.
Table~\ref{tab:obs} summarizes the observations of G335 MM1 undertaken for the DIHCA survey.
The four spectral windows and resolution of the observations are equivalent to those presented in \citetalias{2021ApJ...909..199O}.
To summarize, the spectral windows cover the 216.9--235.5\,GHz range with a spectral resolution of 976.6\,kHz ($\sim$1.3\,km\,s$^{-1}$).

\begin{deluxetable}{lccc}
\tablecaption{ALMA 12\,m observations of G335 MM1 for the DIHCA survey.\label{tab:obs}}
\tablewidth{0pt}
\tablehead{
\colhead{Configuration} & \colhead{Date} & \colhead{Antennas} & \colhead{Baseline} \\
\colhead{}              & \colhead{}     & \colhead{}         & \colhead{range} 
}
\startdata
compact  & November 2016 & 41 & 18.6--1100\,m \\ 
extended & July 2019     & 42 & 92.1--8500\,m \\ 
\enddata
\end{deluxetable}

\begin{figure*}
\begin{center}
\includegraphics[angle=0,scale=0.5]{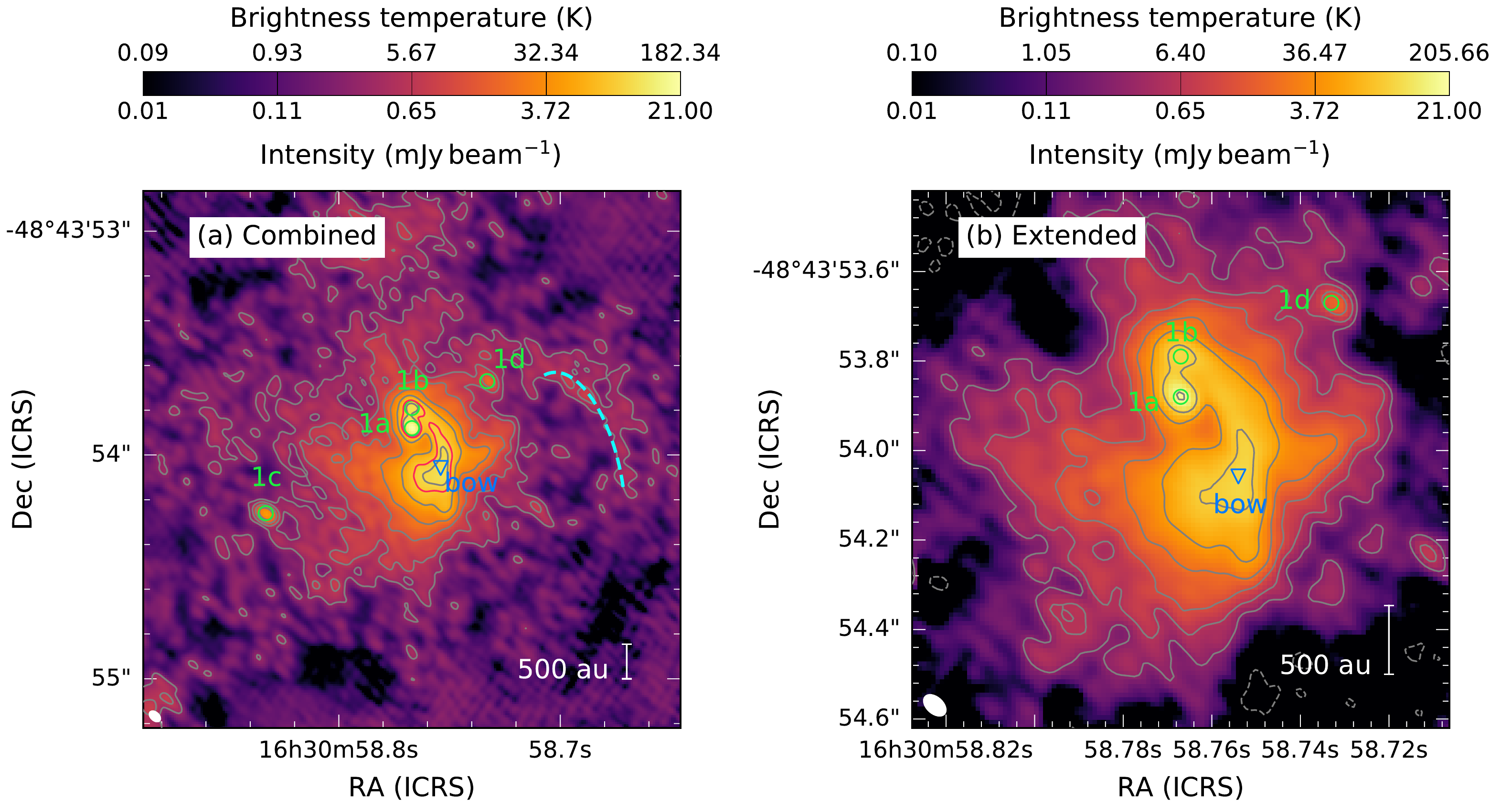}
\end{center}
\caption{ALMA continuum maps of G335 ALMA1 at 1.3\,mm in color scale and contours from (a) the combined dataset and (b) the extended dataset.  
    (a) The grey contours levels are $-3, 5, 10, 20, 40, 80, 160\times\sigma_{\rm cont}$ with $\sigma_{\rm cont}=66$\,$\mu$Jy\,beam$^{-1}$.
    The dashed light blue line shows the arc-shaped structure described in \citetalias{2021ApJ...909..199O}.
    The $130\times\sigma_{\rm cont}$ level is shown in red contours.
(b) The grey contours levels are $-6, -3, 5, 10, 20, 40, 80, 160, 320\times\sigma_{\rm cont}$ with $\sigma_{\rm cont}=57$\,$\mu$Jy\,beam$^{-1}$.
    The green circles and labels mark the peak position of the sources identified within this region and are labeled in order of brightness. 
    The position of the bow structure peak is marked with a blue triangle.
    The beam size is shown in the lower left corner and corresponds to a scale of ${\sim}160$\,au for both data sets.
}
\label{fig:continuum}
\end{figure*}

The data were reduced using CASA \citep[v5.4.0-70;][]{2007ASPC..376..127M}, with J1617-5848, J1427-4206, and J1650-5044 as flux, bandpass, and phase calibrators, respectively.
We then self-calibrated the data and produce continuum maps from line-free channels following the steps detailed in \citetalias{2021ApJ...909..199O}.
In addition to these extended configuration observations, we combined the continuum subtracted visibilities with those from \citetalias{2021ApJ...909..199O} (hereafter compact configuration, see Table~\ref{tab:obs}) to recover large scale diffuse emission.
We produced continuum maps for the extended and combined data sets.
These maps were produced using the TCLEAN task in CASA with Briggs weighting and a robust parameter of 0.5.
The map of the combined data set is shown in Fig.~\ref{fig:continuum}\,(a), and a zoom-in view of the central region from the extended configuration image is shown in Fig.~\ref{fig:continuum}\,(b) for a better contrast of the sources.
The noise level achieved by the continuum extended configuration observations alone is 57\,$\mu$Jy\,beam$^{-1}$, while for the combined data set the noise level is 66\,$\mu$Jy\,beam$^{-1}$. 
The beam FWHM of the continuum CLEAN maps is $0\farcs061\times0\farcs040$ P.A.=48\degr\ and $0\farcs064\times0\farcs043$ P.A.=48\degr\ for the extended and combined continuum images, respectively.
The MRS of the combined data set is 1\farcs74.

Data cubes for lines of interest were produced using the automatic masking procedure YCLEAN \citep{Contreras18}. 
Similar to the continuum, we produced data cubes for the extended and combined data sets.
A noise level of roughly 2\,mJy\,beam$^{-1}$ per channel (${\sim}17$\,K), with a channel width of ${\sim}0.6$\,km\,s$^{-1}$, was achieved in both data sets and across all spectral windows.

Flux measurements in this work are measured in the primary beam corrected images, while figures present uncorrected data.

\section{Results}
\label{sec:results}

\subsection{Core Fragmentation}

The single source ALMA1 resolved by previous 0\farcs3 resolution ALMA observations further divides into at least 4 continuum sources identified in the combined map.
Figure~\ref{fig:continuum}\,(a) shows the continuum observations with the new sources labeled.
Table~\ref{tab:props} lists the properties of the sources measured from a 2-D Gaussian fit to the continuum emission. 
We use the combined data set to measure the source fluxes. 
The brightest source (ALMA1a) has a close companion (ALMA1b) separated by ${\sim}85$\,mas, corresponding to a projected distance of ${\sim}280$\,au.
The two additional sources (ALMA1c and ALMA1d) are located at a projected distance of 2500 and 1300\,au from ALMA1a, respectively.

\begin{deluxetable*}{lccccccc}
\tablecaption{Continuum Source Properties\label{tab:props}}
\tablewidth{0pt}
\tablehead{
\colhead{ALMA} & \colhead{R.A. (ICRS)} & \colhead{Decl. (ICRS)} & \colhead{$I_{\rm peak}$} & \colhead{$T_{b}$} & \colhead{$F_{\rm total}$} & \colhead{FWHM} & \colhead{$v_{\rm LSR}$}\\
\colhead{Source} & \colhead{$[h:m:s]$} & \colhead{$[\degr:\arcmin:\arcsec]$} & \colhead{(mJy\,beam$^{-1}$)} & \colhead{(K)} & \colhead{(mJy)} & \colhead{(mas)} & \colhead{(km s$^{-1}$)}
}
\startdata
1a  & 16:30:58.767 & -48:43:53.87 & 20.6  & 179 & $69\pm6$      & $101\times80$  & --46.9 \\ 
1b  & 16:30:58.767 & -48:43:53.80 & 11.8  & 103 & $72\pm3$      & $120\times115$ & --46.9 \\ 
1c  & 16:30:58.833 & -48:43:54.26 &  4.7  &  41 & $7.2\pm0.4$   &  $47\times35$  & --46.9 \\ 
1d  & 16:30:58.733 & -48:43:53.67 &  2.1  &  18 & $6.8\pm1.1$   & $103\times90$  & [--56.5,--53.2] \\ 
bow\tablenotemark{a} & \nodata  & \nodata & 12.1 & 105 & 93  & \nodata & \nodata \\
\enddata
\tablecomments{Fluxes measured on the combined primary beam corrected data set. The FWHM is deconvolved from the beam.}
\tablenotetext{a}{Fluxes measured within the $130\sigma$ contour with $\sigma=66$\,$\mu$Jy\,beam$^{-1}$. This is an arbitrary number defined to trace most of the bow emission without including ALMA1a and ALMA1b.}
\end{deluxetable*}

An additional structure is observed to the south of ALMA1a at a projected distance of ${\sim}$780\,au (0\farcs24 measured peak to peak).
This bow-shaped object is connected to the central region by an arc-shaped structure.
This source is labeled as ``bow'' in Figure~\ref{fig:continuum} and its properties are listed in Table~\ref{tab:props}.
We avoid categorizing this structure as a core because of its shape and location,  which is close to the large scale molecular outflow \citepalias[P.A.=210\degr;][]{2021ApJ...909..199O}. 
We discuss different scenarios for the origin of this source in Section~\ref{sec:discussion}.
In order to facilitate the analysis, we will refer to the region enclosing the sources ALMA1a, ALMA1b and bow as the central region, as shown in Figure~\ref{fig:continuum}(b).
This region encompasses ${\sim}70$\% of the flux density of ALMA1 ($F_{\rm ALMA1}=0.74$\,Jy).
The remaining ${\sim}30$\% is produced by ALMA1c and extended/fainter structures, e.g., the arc-shaped structure in Figure~\ref{fig:continuum}(a).

\subsection{Line emission}

\begin{figure*}
\begin{center}
\includegraphics[scale=0.38]{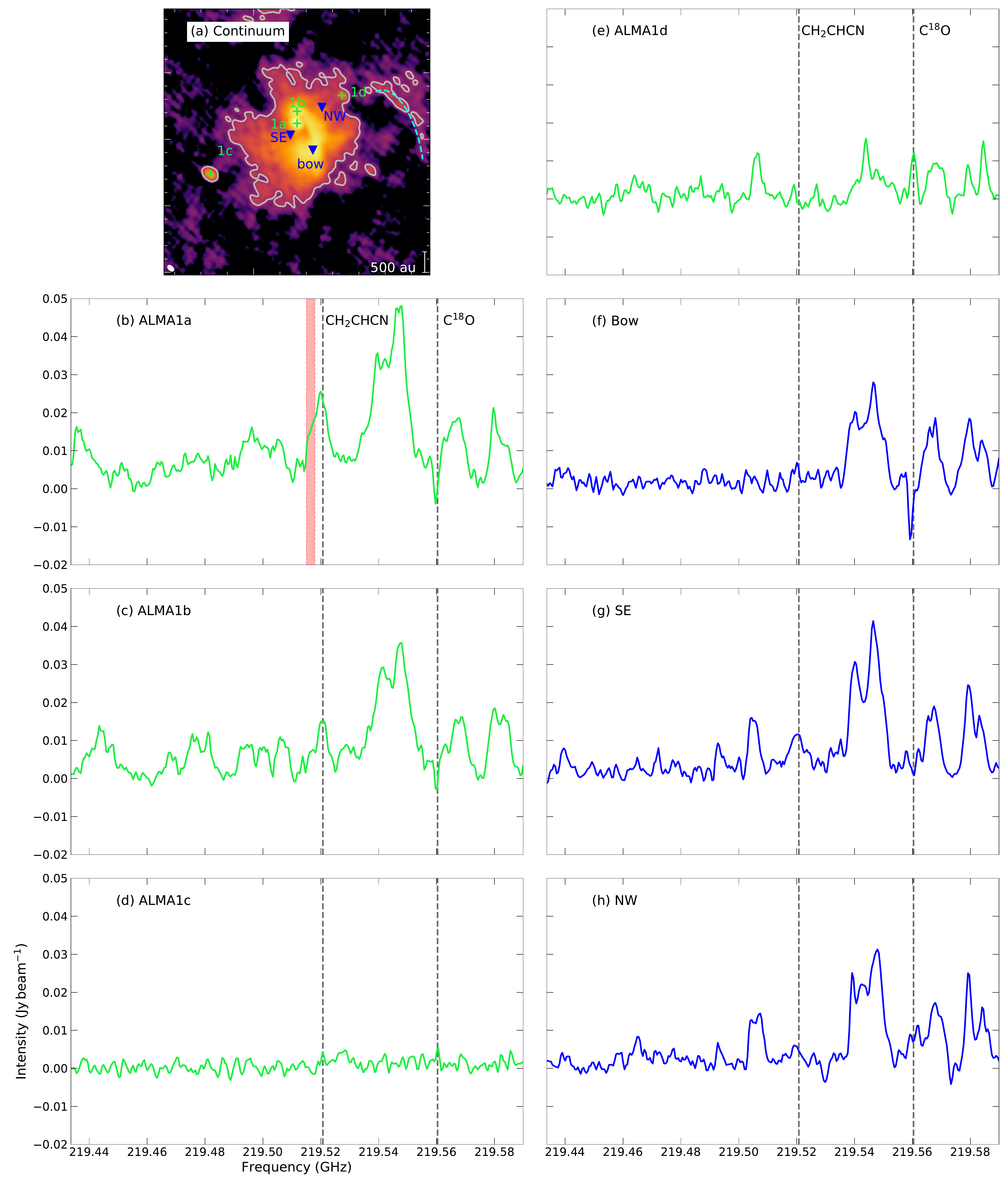}
\end{center}
\caption{Example spectra toward selected positions marked in the dust continuum image displayed in panel (a).
The red shaded area highlights the high-velocity red wing of the CH$_2$CHCN line (see Section~\ref{sec:results:kin}).
}
\label{fig:lines}
\end{figure*}

Table~\ref{tab:lines} summarizes the molecules and transitions analyzed here, and the type of line profile observed towards the brightest continuum source.
As in the compact configuration observations, several molecular line transitions are observed in general.
However, toward sources ALMA1a and ALMA1b we observe self-absorbed transitions of, e.g., H$_2$CO, and inverse P-Cygni profiles of, e.g., CH$_3$CN (see Section~\ref{sec:results:kin}).
Figure~\ref{fig:lines} shows example spectra towards different positions in one of the spectral windows.
The bow is devoid of molecular lines for the most part, and molecules tracing cold, lower density gas, like CO isotopologues, are observed in absorption as shown in Figure~\ref{fig:lines}(f) for C$^{18}$O. 
It is worth noticing that many lines start to disappear toward the continuum peaks when compared to the compact configuration data from Paper I.

\begin{deluxetable*}{lccccc}
\tablecaption{Molecular Lines and Transitions Analyzed.\label{tab:lines}}
\tablewidth{0pt}
\tablehead{
    \colhead{Molecule} & \colhead{Transition} & \colhead{Freq.}  & \colhead{$E_u$} & \colhead{Line} & \colhead{Ref.} \\
    \colhead{}         & \colhead{}           & \colhead{(GHz)}  & \colhead{(K)}   &  \colhead{Profile} &\colhead{}
}
\startdata
SiO         & $J=5-4$         & 217.1049800 & 31 & self-absorbed   & (2) \\
SO          & $^3\Sigma$ $v=0$ $J_K=6_5-5_4$ & 219.949442 & 35 & self-absorbed & (1) \\
$^{13}$CO   & $J=2-1$         & 220.4758072 & 16 & inverse P-Cygni & (2) \\ 
CH$_3$CN    & $J_K=12_3-11_3$ & 220.7090165 & 133 & inverse P-Cygni & (1) \\ 
CH$_2$CHCN  & $v_{11}=1$ $J_{K_a,K_c}=23_{15,8}-22_{15,7}$ & 219.5207483 & 929 & single & (1) \\ 
CH$_3$OH    & E$_1$ $v_t=0$ $J_{K}=25_{3}-24_{4}$ & 219.983675 & 802 & single & (2)\\
            & E$_1$ $v_t=0$ $J_{K}=23_{5}-22_{6}$ & 219.993658 & 776 & single & (2)\\
\enddata
\tablerefs{
(1) Jet Propulsion Laboratory \citep[JPL,][]{1998JQSRT..60..883P};
(2) Cologne Database for Molecular Spectroscopy \citep[CDMS,][]{2005JMoSt.742..215M}.
}
\end{deluxetable*}

We detect faint emission that may be associated with blue-shifted H30$\alpha$ emission ($\nu=231.90092784$\,GHz). 
Alternative lines in the same frequency range (231.8986396--231.9035225\,GHz) include transitions of $^{33}$SO$_2$ with the one with the highest Einstein coefficient at 231.9002488\,GHz (CDMS; $\left| \Delta \nu_{{\rm H}30\alpha}\right| =0.7$\,MHz), CH$_3$C$^{15}$CN at 231.90223\,GHz (JPL; $\left| \Delta \nu_{{\rm H}30\alpha}\right| =1.3$\,MHz) and other carbon bearing molecules.
Appendix Figure~\ref{fig:ap:halpha} shows the zeroth and first order moment maps.
The emission is distributed in the direction of the blue-shifted outflow lobe surrounding the sources ALMA1a, ALMA1b and bow.
The emission toward these sources is attenuated or extinct by the optically thicker continuum (see below).
The diameter of the emission is roughly 2600\,au (0.01\,pc), which is larger than the size of the \ion{H}{2} region estimated by \citet[][${\sim}95$\,au]{2015A&A...577A..30A}.
Given its distribution, velocity shift and extent, the H30$\alpha$ emission may be associated with gas ionized by photons escaping through the less dense medium of the outflow cavity.

\subsection{Physical Properties}
\label{sec:results:props}

\begin{figure}
\begin{center}
\includegraphics[scale=0.48]{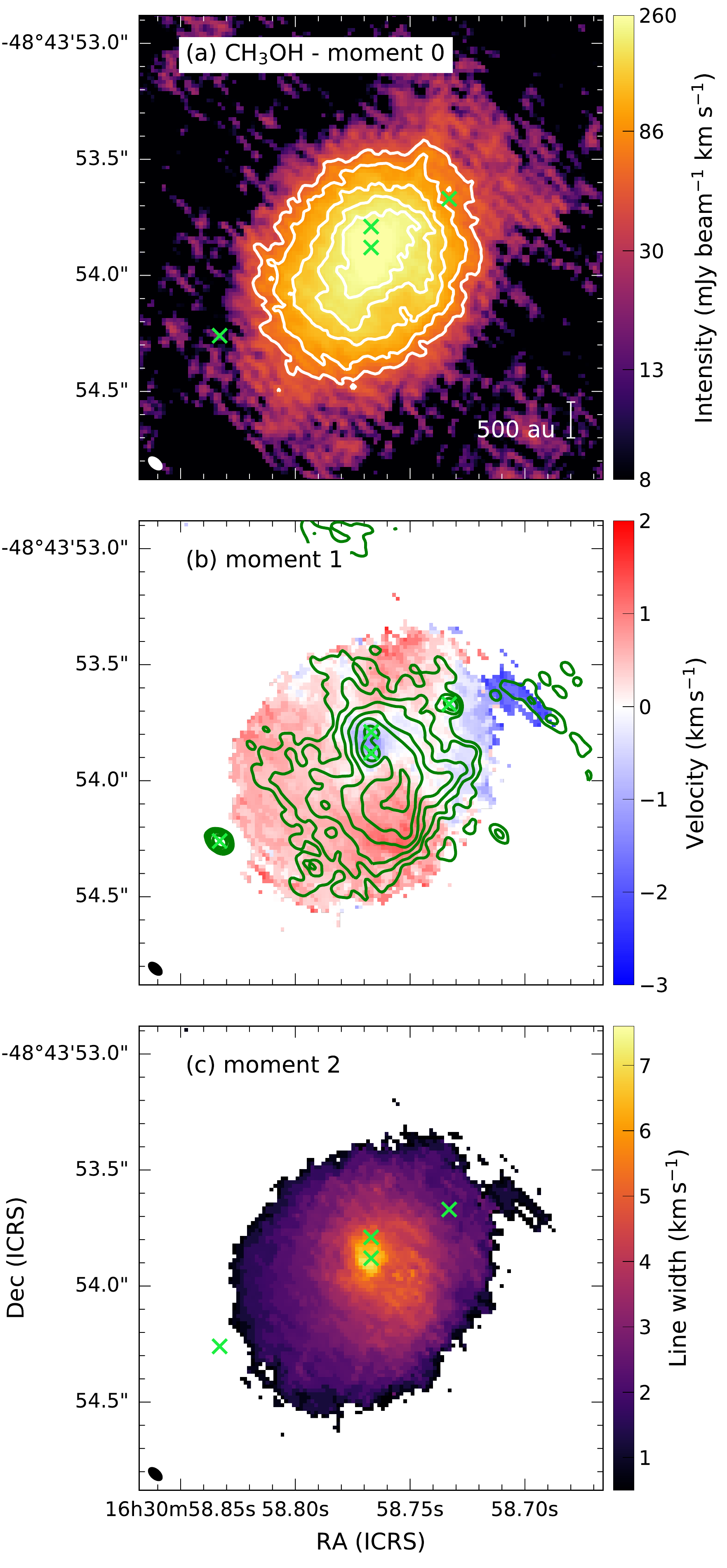}
\end{center}
\caption{CH$_3$OH $J_{K}=23_{5}-22_{6}$ moment maps from the combined data set. 
(a) Zeroth order moment map.
Contour levels correspond to 5, 8, 12, 16, $20\times \sigma_{\rm rms}$ with $\sigma_{\rm rms}=11$\,mJy\,beam$^{-1}$\,km\,s$^{-1}$.
(b) First moment map with continuum emission in contours from Figure~\ref{fig:continuum}(b).
Zero systemic velocity corresponds to the source $v_{\rm LSR}$ (--46.9\,km\,s$^{-1}$).
(c) Second order moment map.
The location of the continuum sources are marked with green crosses.
The beam size is shown in the lower left corner.
}
\label{fig:methanol}
\end{figure}

Figure~\ref{fig:methanol} shows the moment maps of the CH$_3$OH $J_{K}=23_{5}-22_{6}$ transition listed in Table~\ref{tab:lines} as an example.
The moments are calculated in a spectral window with a width of 20 channels (${\sim}12$\,km\,s$^{-1}$) centered at the line frequency, and first and second moments are calculated from emission over $5\sigma$ with $\sigma=2$\,mJy\,beam$^{-1}$.
Figure~\ref{fig:methanol}(a) shows that the emission is extended, tracing gas around the central region of ALMA1 (including 1a, 1b, and the bow), in what seems to be a common reservoir of gas. 
As such, we use the CH$_3$OH transitions listed in Table~\ref{tab:lines} to derive  physical properties, namely the circumstellar gas temperature, CH$_3$OH column density, velocity distribution, and line width. 
We fitted the spectra on a pixel-by-pixel basis using the CASSIS software\footnote{ CASSIS has been developed by IRAP-UPS/CNRS (\url{http://cassis.irap.omp.eu}).} \citep{2015sf2a.conf..313V} and the CDMS molecular database \citep{2005JMoSt.742..215M}.
The fitting assumes local thermodynamic equilibrium (LTE) in a column of gas with constant density, and a Gaussian line shape.
An additional parameter is required for the fit, the source size, to determine the beam dilution factor.
We set the source size to 1\arcsec, which is roughly the size of the CH$_3$OH emission, because the source is well resolved. 
We limit the fit only to data over the $5\sigma$ level in the zeroth order moment map (see Figure~\ref{fig:methanol}a).
We use the Markov Chain Monte Carlo tasks built in CASSIS to explore the parameter space.
In general, the lines are well fitted with a single temperature/density component, resulting in reduced $\chi^2$ values below 2.

\begin{figure*}
\begin{center}
\includegraphics[scale=0.48]{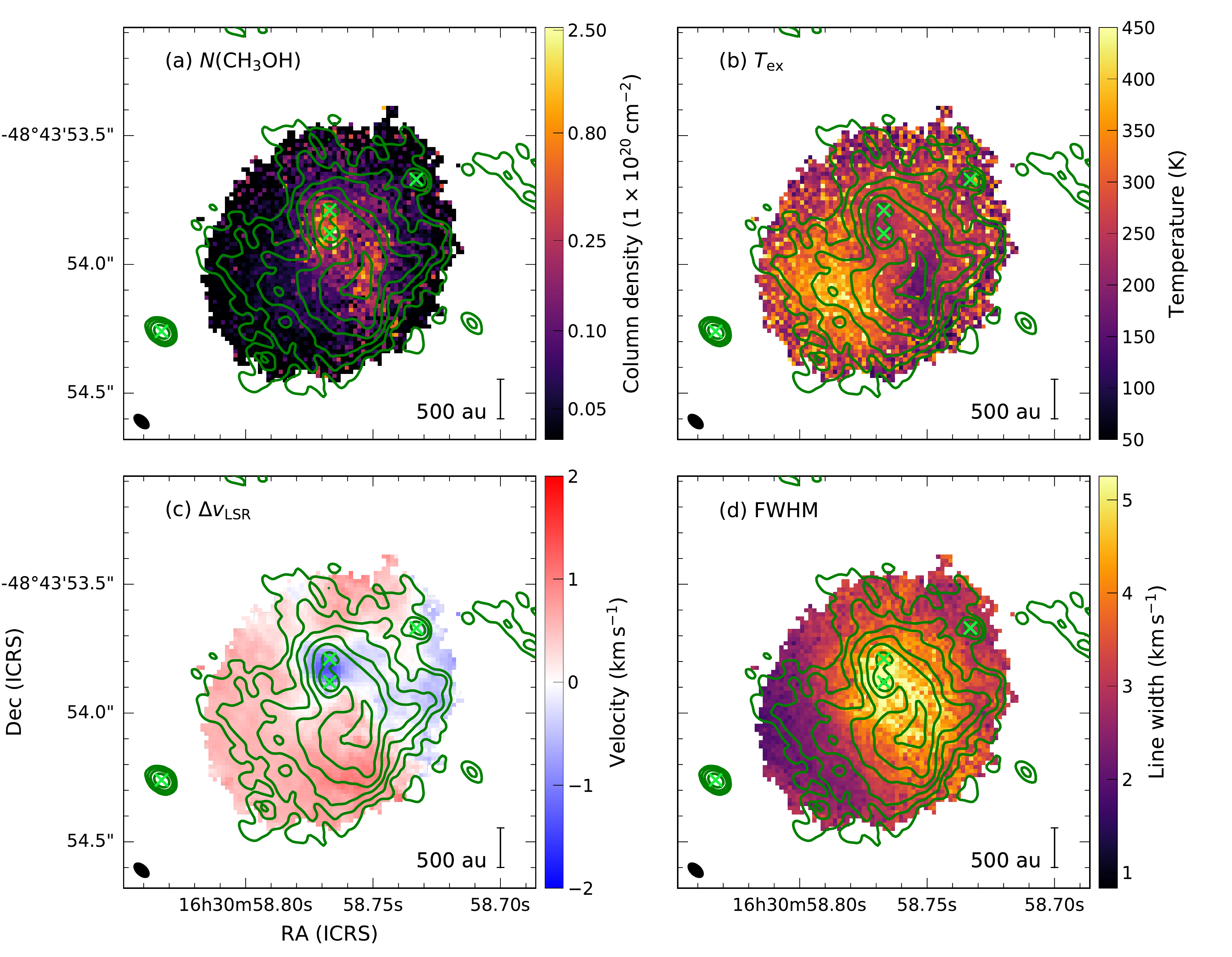}
\end{center}
\caption{CH$_3$OH LTE modeling results.
(a) CH$_3$OH column density.
(b) Excitation temperature map.
(c) Velocity shift with respect to the systemic velocity.
Zero systemic velocity corresponds to the source $v_{\rm LSR}$ (--46.9\,km\,s$^{-1}$).
(d) Line FWHM map.
Contour levels correspond to 5, 8, 12, 16, $20\times \sigma_{\rm rms}$ with $\sigma_{\rm rms}=11$\,mJy\,beam$^{-1}$.
The green contours correspond to the continuum emission from the extended configuration, Figure~\ref{fig:continuum}(b). 
The location of the continuum sources are marked with green crosses.
The beam size is shown in the lower left corner. 
}
\label{fig:methanol:cassis}
\end{figure*}

Figure~\ref{fig:methanol:cassis} shows the distribution of the fitted physical properties, while the error maps for these properties and the reduced $\chi^2$ map from the fit are presented in the Appendix Figure~\ref{fig:ap:methanol:cassis:error}.
The fitted spectra towards the continuum peak positions of ALMA1a, ALMA1b and bow are presented in the Appendix Figure~\ref{fig:ap:methanol}.
The column density peaks are located at the position of sources ALMA1a and ALMA1b, and towards the bow structure, with average values over a beam-sized region between $2-9\times10^{19}$\,cm$^{-2}$.
Note that the errors in column density are particularly large in the region surrounding the sources, this is likely due to the number of lines fitted and a lower S/N.
Similarly, the average temperatures around these sources are $\sim220$\,K.
Table~\ref{tab:phys:props} lists the column density and temperature values for each source and the median values of the whole ALMA1 region.
We note, however, that higher temperatures are achieved around the sources.
The velocity distribution and line width maps in Figures~\ref{fig:methanol:cassis}(c) and (d) are consistent with the first and second moment maps in Figures~\ref{fig:methanol}(b) and (c), respectively.

\begin{deluxetable*}{lcccccc}
\tablecaption{Source Physical Properties\label{tab:phys:props}}
\tablewidth{0pt}
\tablehead{
\colhead{ALMA} & \colhead{$R$\tablenotemark{a}} & \colhead{$T$} & \colhead{$L$} & \colhead{$M_{d}$} & \colhead{$N^{\rm peak}_{\rm H_2}$} & \colhead{$N_{\rm CH_3OH}$}\\
\colhead{Source} & \colhead{(au)} & \colhead{(K)} & \colhead{(\lsun)} & \colhead{(\msun)} & \colhead{($10^{25}$\,cm$^{-2}$)} & \colhead{($10^{19}$\,cm$^{-2}$)}
}
\startdata
\cutinhead{Optically thin}
1\tablenotemark{b} & 710 & 265 & \nodata & 7.0 & 0.4 & 0.5 \\
1a  & 149 & $230\pm9$& \nodata & $>1.0$ & $>1.7$ & $8.4\pm3.6$    \\ 
1b  & 192 & $220\pm5$ & \nodata & $>1.1$ & $>1.0$ & $5.7\pm2.0$     \\ 
1c  & 66  & 20  & \nodata & 1.5 & 5.8 & \nodata \\ 
1d  & 157 & 20  & \nodata & 1.5 & 2.6 & \nodata \\ 
bow & 300 & $220\pm2$  & \nodata & $>1.4$  & $>1.0$ & $2.5\pm0.2$ \\
\cutinhead{Optically thick}
1a   &  149    & 179 & $>950$ & \nodata & \nodata & \nodata \\
1b   &  192    & 103 & $>170$ & \nodata & \nodata & \nodata \\ 
bow  &  300    & 105 & $>480$ & \nodata & \nodata & \nodata \\
\enddata
\tablecomments{
The temperature and CH$_3$OH column density of ALMA1a and ALMA1b are averages over a beam sized area with a radius of 26\,mas from the maps in Figure~\ref{fig:methanol:cassis}, while their errors are calculated using $\sigma=N^{-1}\sqrt{\sum \sigma_i^2}$ with the $N$ corresponding values of $\sigma_i$ in Appendix Figure~\ref{fig:ap:methanol:cassis:error}.
The same properties for the bow structure are measured over the same region as the measured fluxes in Table~\ref{tab:props} ($130\sigma$ level).}
\tablenotetext{a}{The radius of ALMA1 is from \citetalias{2021ApJ...909..199O}, while radius of sources 1a--1d corresponds to half of the geometric mean of the deconvolved sizes (FWHM) from Table~\ref{tab:props}.
The radius of the bow is determined by the area of the $130\sigma$ level with $A_{130\sigma}=\pi r^2$.
This is an arbitrary value as the source is non-Gaussian.}
\tablenotetext{b}{The temperature and CH$_3$OH column density correspond to the median values over the ALMA1 region from the maps in Figure~\ref{fig:methanol:cassis}. The mass and peak column density values are calculated from data in \citetalias{2021ApJ...909..199O}: $F_{\rm total}=566$\,mJy and $I_{\rm peak}=209.8$\,mJy\,beam$^{-1}$.}
\end{deluxetable*}

Following \citetalias{2021ApJ...909..199O} and assuming optically thin dust emission, we calculate the gas mass as
\begin{equation}\label{eq:dustmass}
    M_d = \frac{F_\nu d^2 R_{gd}}{\kappa_\nu B_\nu(T_d)}
\end{equation}
with $F_{1.3{\rm mm}}$ the flux density from Table~\ref{tab:props}, $d=3.25$\,kpc the source distance, $R_{\rm gd}=100$ the gas-to-dust mass ratio, $\kappa_{1.3{\rm mm}}=1$\,cm$^2$\,gr$^{-1}$ the dust opacity \citep{1994A&A...291..943O}, and $B_\nu$ the Planck blackbody function.
We also calculate the peak H$_2$ column density, defined as
\begin{equation}
  N^{\rm peak}_{{\rm H}_2} = \frac{I_\nu R_{gd}}{B_\nu(T_d) \kappa_\nu \mu_{\rm H_2} m_{\rm H}}
\end{equation}
with $I_{1.3{\rm mm}}$ the peak intensity from Table~\ref{tab:props}, $\mu_{\rm H_2}=2.8$ the molecular weight per hydrogen molecule \citep[e.g.,][]{2008A&A...487..993K}, and $m_{\rm H}$ the atomic hydrogen mass.
The lack of line emission makes difficult an accurate estimation of the dust temperature for all sources, hence we have adopted different approaches to estimate the dust temperature.
For sources ALMA1a and ALMA1b we use the temperature estimates from the fit  to the CH$_3$OH emission.
Sources ALMA1c and ALMA1d likely host low-mass protostars, thus we use a dust temperature of 20\,K. 
The mass and column density of the sources are listed in Table~\ref{tab:phys:props} (optically thin heading).

Note that the optical depth $\tau_\nu = \mu_{\rm H_2} m_{\rm H} \kappa_\nu R_{gd}^{-1} N^{\rm peak}_{\rm H_2}$=1 when $N^{\rm peak}_{\rm H_2}=2.1\times10^{25}$\,cm$^{-2}$.
Thus for sources ALMA1a, ALMA1b and the bow structure the dust emission is becoming optically thick.
This is also true for the column density derived from CH$_3$OH for abundances as high as $10^{-6}$ \citep[e.g.,][]{1986A&A...157..318M,2014ApJ...787..112C}.
We thus consider the values of the column densities for sources ALMA1a, ALMA1b and bow as lower limits.
For sources ALMA1c and ALMA1d the temperatures may be higher, in which case the column densities are an upper limit.

\begin{figure}
\begin{center}
\includegraphics[scale=0.75]{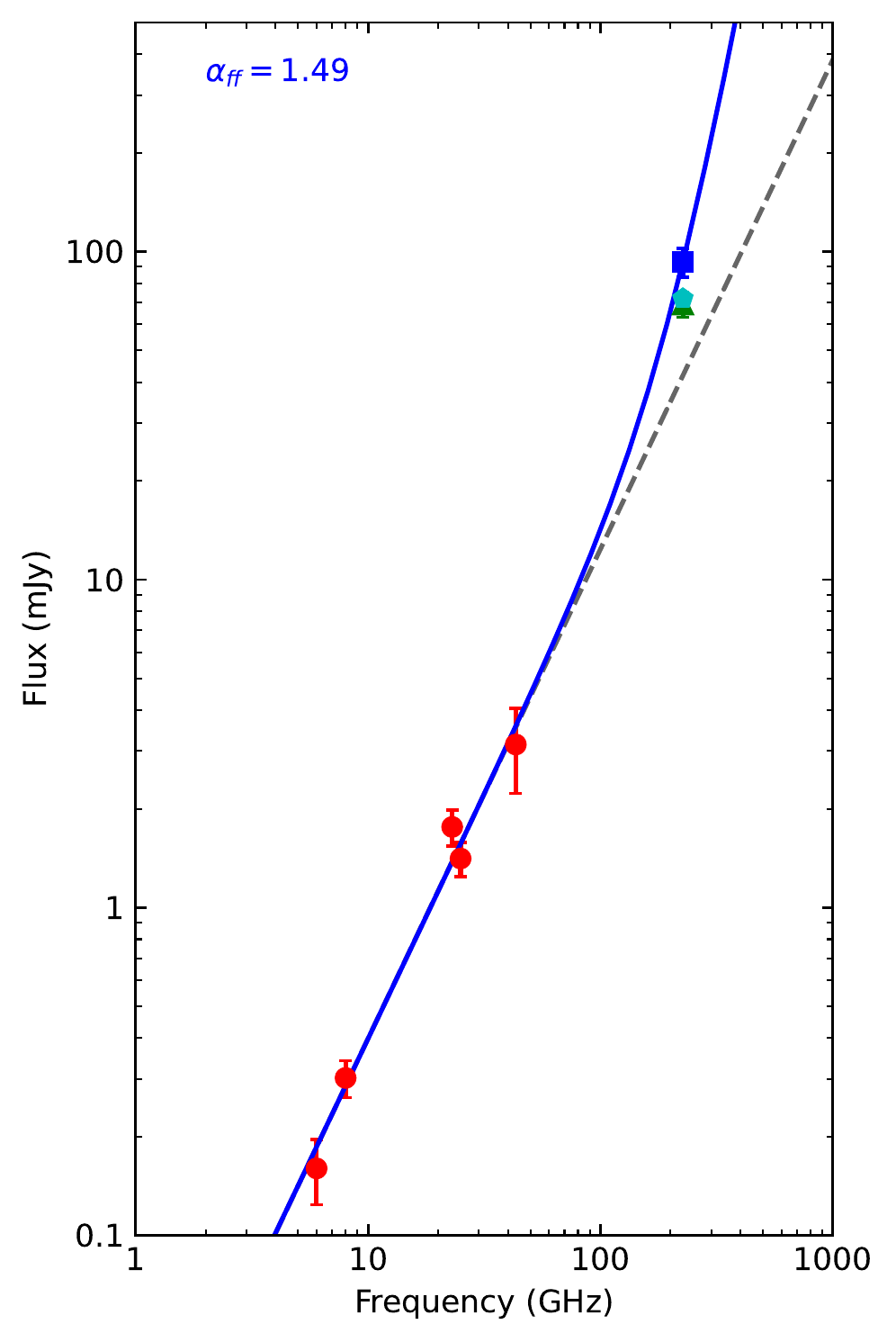}
\end{center}
\caption{Radio and mm SED of G335 ALMA1.
Blue square is the contribution from the bow structure (see Table~\ref{tab:props}).
Red dots correspond to the radio measurements of G335 MM1a from \citet{2015A&A...577A..30A}.
The green triangle and  cyan pentagon correspond to the contributions from ALMA1a and 1b, respectively.
The blue line corresponds to a model with two power law distributions accounting for  the contributions of dust and free-free and assuming that the radio free-free is produced by the bow structure.
The dashed line corresponds to a power law fitted to the radio data.
}
\label{fig:sed}
\end{figure}

Similarly, the masses of ALMA1a, ALMA1b and the bow structure are also a lower limit.
These sources are located at the center of the gravitational potential well of ALMA1 (6-19\,\msun, see \citetalias[][]{2021ApJ...909..199O} and Table~\ref{tab:phys:props}) as shown in the Appendix Figure~\ref{fig:ap:alma1} and the larger MM1 clump \citep[790\,\msun;][see also their Figure 3]{2015A&A...577A..30A}.
Hence, the gas reservoir around the continuum sources can be replenished and the forming stars can continue growing following scenarios like competitive accretion \citep[e.g.,][]{2009MNRAS.400.1775S} or global hierarchical collapse \citep[][]{2019MNRAS.490.3061V}, see also discussions in \cite{Contreras18} and \cite{Sanhueza19}. 
The masses derived under the optically thin approximation although lower limits are similar to those found in other high-mass binary systems at equivalent radius (e.g., 0.2 and 0.04\,\msun\ at 150\,au for the detected objects in IRAS 16547--4247, \citealp{2020ApJ...900L...2T}).
Contribution from free-free also becomes important for the mass and column density calculations at the scales of the observations presented here.
Following the fitting of the free-free and dust continuum emission of \citet{2015A&A...577A..30A} with a turnover frequency of ${\sim}22$\,GHz, a contribution from free-free emission of less than 5\,mJy was estimated in \citetalias{2021ApJ...909..199O}, which is roughly an order of magnitude lower than the flux density of the brightest source.
However, HC \ion{H}{2} regions may have higher turnover frequencies if a density gradient is present \citep[][and references therein]{2008ASPC..387..232L}.
Figure~\ref{fig:sed} shows the spectral energy distribution with a power law fit to the radio data.
Assuming the radio emission is only due to free-free, we expect its contribution at 220\,GHz to be 42\,mJy, which is roughly a 60\% of the continuum emission of ALMA1a at the same frequency and hence non-negligible.
Nevertheless, the tentative H30$\alpha$ emission in the Appendix Figure~\ref{fig:ap:halpha} indicates that the free-free contribution may be more extended, hence contributing less to the continuum of each individual mm source.

In the optically thick limit, the thermal radiation can be approximated by black body radiation and the dust brightness temperature, $T_b$, converges to the dust temperature.
Table~\ref{tab:props} lists the values of the brightness temperature.
Only the values of the brightness temperature and the temperature from CH$_3$OH for ALMA1a are relatively similar.
We estimate the luminosity of the sources by using 
\begin{equation}
    L= 4\pi r^{2} \sigma_{sb} T^4
\end{equation}
with $r$ the source radius and $\sigma_{sb}$ the Stefan-Boltzmann constant.
The values of the luminosity for $T=T_b$ are listed in Table~\ref{tab:phys:props} (optically thick heading).
These values are likely lower limits as the beam filling factor may be different lower than unity, due to clumpiness within the source. 
Based on the disk accretion models of \citet{2010ApJ...721..478H}, at the estimated luminosity the high-mass (proto-)stars would have accumulated masses in excess of 4-5\,\msun, and would be on or have finished the bloated stage where the luminosity sharply increases depending on the accretion rate.

\subsection{Kinematics}
\label{sec:results:kin}

The large scale emission from CH$_3$OH is likely produced by a combination of motions.
Towards the blue-shifted outflow lobe (south-west), we see blue-shifted emission and wider line widths as shown by both the observations and model (Figures~\ref{fig:methanol} and \ref{fig:methanol:cassis}).
This pattern was also observed in the high-mass YSO AFGL-2591, and interpreted as gas in the surface of the envelope cavity walls being shocked by the outflow \citep{2020MNRAS.498.4721O}.
Elsewhere, the velocity pattern shows a central blue-shifted region surrounded by red-shifted emission.
Upon inspection of the spectra between the sources ALMA1a and 1b, we note that the spectra have blue wings that produce the blue spot at the center, rather than blue skewed profiles indicative of infall (e.g., \citealp{2019A&A...626A..84E}, \citetalias{2021ApJ...909..199O}).
These can be produced by the motions of two or more gas components, e.g., the combined effect of the rotation of the cores, as well as the blue-shifted lobe of the outflow closer to the source.

\begin{figure*}
\begin{center}
\includegraphics[scale=0.48]{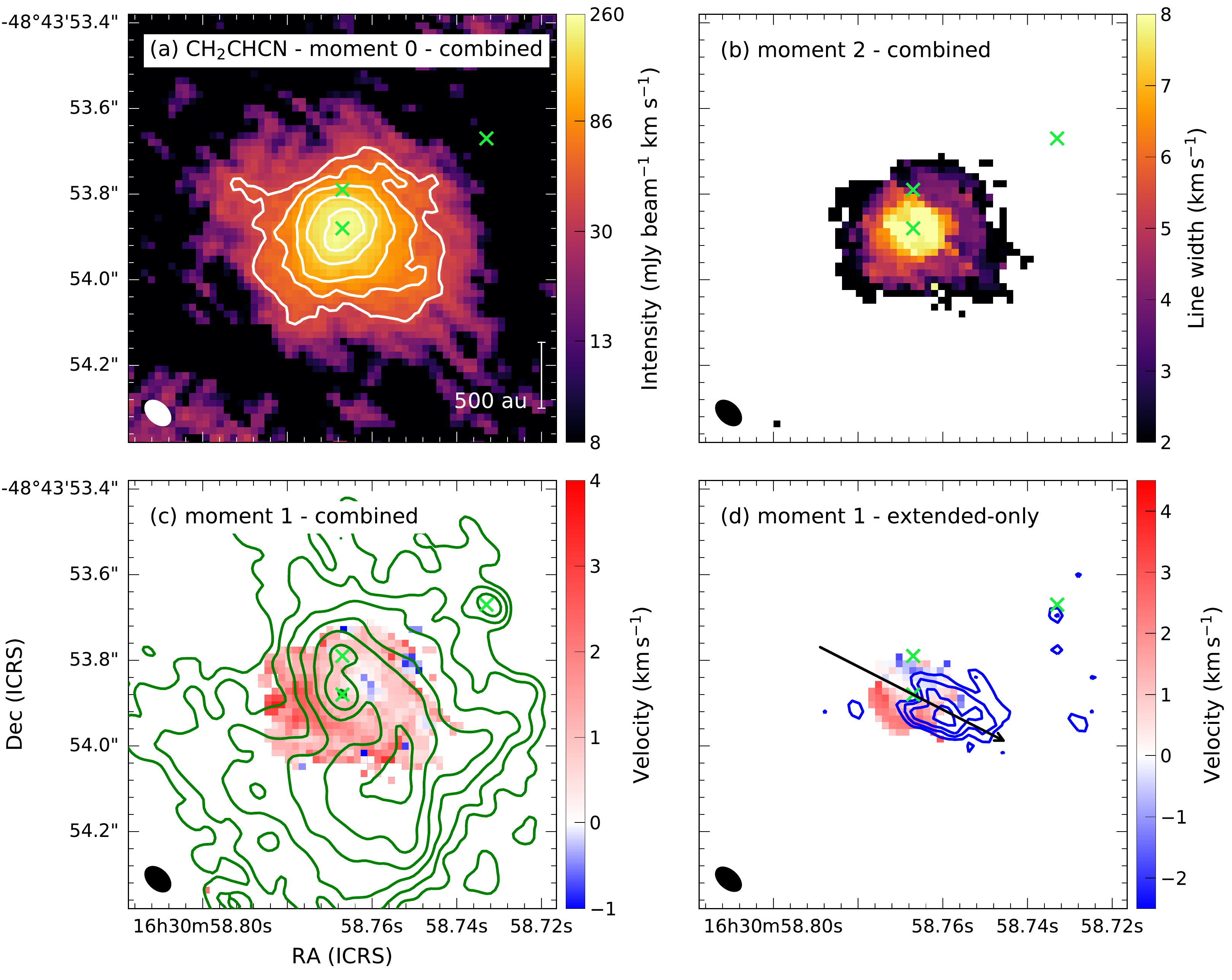}
\end{center}
\caption{CH$_2$CHCN moment maps.
(a) Zeroth moment map.
Contour levels are 5, 8, 12, 18, $24\times\sigma_{\rm rms}$ with $\sigma_{\rm rms}=8.6$\,mJy\,beam$^{-1}$\,km\,s$^{-1}$.
(b) Second moment map of the emission over the $5\sigma$ level with $\sigma=2$\,mJy\,beam$^{-1}$ the noise level per channel.
(c) Fist moment map of the emission over $5\sigma$ level from the combined data set.
The green contours correspond to the continuum emission from Figure~\ref{fig:continuum}(b).
Zero systemic velocity corresponds to the source $v_{\rm LSR}$ (--46.9\,km\,s$^{-1}$).
(d) First moment map from the extended configuration data set.
The arrow indicates the direction perpendicular to velocity gradient (P.A.=150\degr), i.e. along the rotation axis.
Blue contours correspond to the blue-shifted outflow lobe emission from SiO (see Figure~\ref{fig:sio:split}).
The locations of the continuum sources are marked with green crosses.
The beam size is shown in the lower left corner. 
}
\label{fig:vinyl}
\end{figure*}

Among the myriad of lines in the spectrum we could find only one single-peaked line tracing the region close to the continuum sources.
The line emission is likely produced by a transition of CH$_2$CHCN (vinyl cyanide), and traces the inner region of ALMA1a. We note that vinyl cyanide has been previously detected in extreme high-mass star formation environments, such as Orion KL \citep{Lopez14} and Sgr B2(N) \citep{Belloche13}.
Figure~\ref{fig:vinyl} shows the moment maps from CH$_2$CHCN.
The zeroth order moment map in Figure~\ref{fig:vinyl}a peaks at the same position of ALMA1a and shows that the line is tracing the inner 500\,au of the source.
The velocity distribution in the first moment map from the emission traced by the combined observations (Figure~\ref{fig:vinyl}c) is mostly shifted towards the red.
This is probably produced by high-velocity red wings (see red shaded area in Figure~\ref{fig:lines}b) similar to those observed in HDCO in \citetalias{2021ApJ...909..199O}.
On the other hand, the first moment map from the extended configuration data (Figure~\ref{fig:vinyl}d) shows a velocity gradient in the north-west to south-east direction, perpendicular to the molecular outflow. 
We measure a velocity gradient with a ${\rm P.A.} = 150\degr\pm28\degr$ in a region about the size of the beam at the continuum source position.

From the velocity gradient we also estimate the mass of the ALMA1a source inside a radius $r$.
We measure a velocity difference between the blue- and red-shifted lobes of the first moment map in Figure~\ref{fig:vinyl}d of 3\,km\,s$^{-1}$ at points separated by 0\farcs14 (455\,au) in a slit passing through the source position and with a P.A. equal to that of the velocity gradient.
The velocity along the line of sight is thus $|v_{\phi,obs}|=1.5$\,km\,s$^{-1}$ at a distance of $r=228$\,au from the source.
For a purely rotating disk, the velocity of rotation is related to the observed velocity by $|v_{\phi,obs}|=|v_{\phi}|\sin i$.
Assuming Keplerian rotation (see position velocity, pv, map in Appendix Figure~\ref{fig:ap:pvmap}) the mass inside $r$ is given by
\begin{equation}
    M = \frac{r v_{\phi,obs}^2}{G \sin^2 i}~,
\end{equation}
with $G$ the gravitational constant.
The deconvolved FWHM of the CH$_2$CHCN emission from a 2-D Gaussian fit to the zeroth order moment maps are $0\farcs281\times0\farcs252$ and $0\farcs162\times0\farcs0.145$ for the combined and extended configuration data, respectively.
The inclination angle can be estimated by assuming that the emission comes from a disk with $\cos i = b/a$, where $a$ and $b$ the semi-major and -minor axes.
We obtain a mean inclination angle of $26\degr$, which is lower than those estimated by \citep[][$i=57$\degr\ and 76\degr]{2021A&A...645A.142A}.
This change in inclination can be the result of precession caused by the binary system.
Moreover, the change in orientation of the SiO outflow emission between the small and large scales (see below) indicates that we may be looking into the outflow cavity (lower inclination angle).
We thus obtain a star+disk mass of 3\,\msun.
For a B1-1.5 zero-age main sequence star \citep{2015A&A...577A..30A} with a mass of ${\sim}7$\,\msun, the inclination angle would need to be $i{\sim}15\degr$ to explain the observed velocity distribution.

\subsubsection{Outflows}

\begin{figure}
\begin{center}
\includegraphics[scale=0.48]{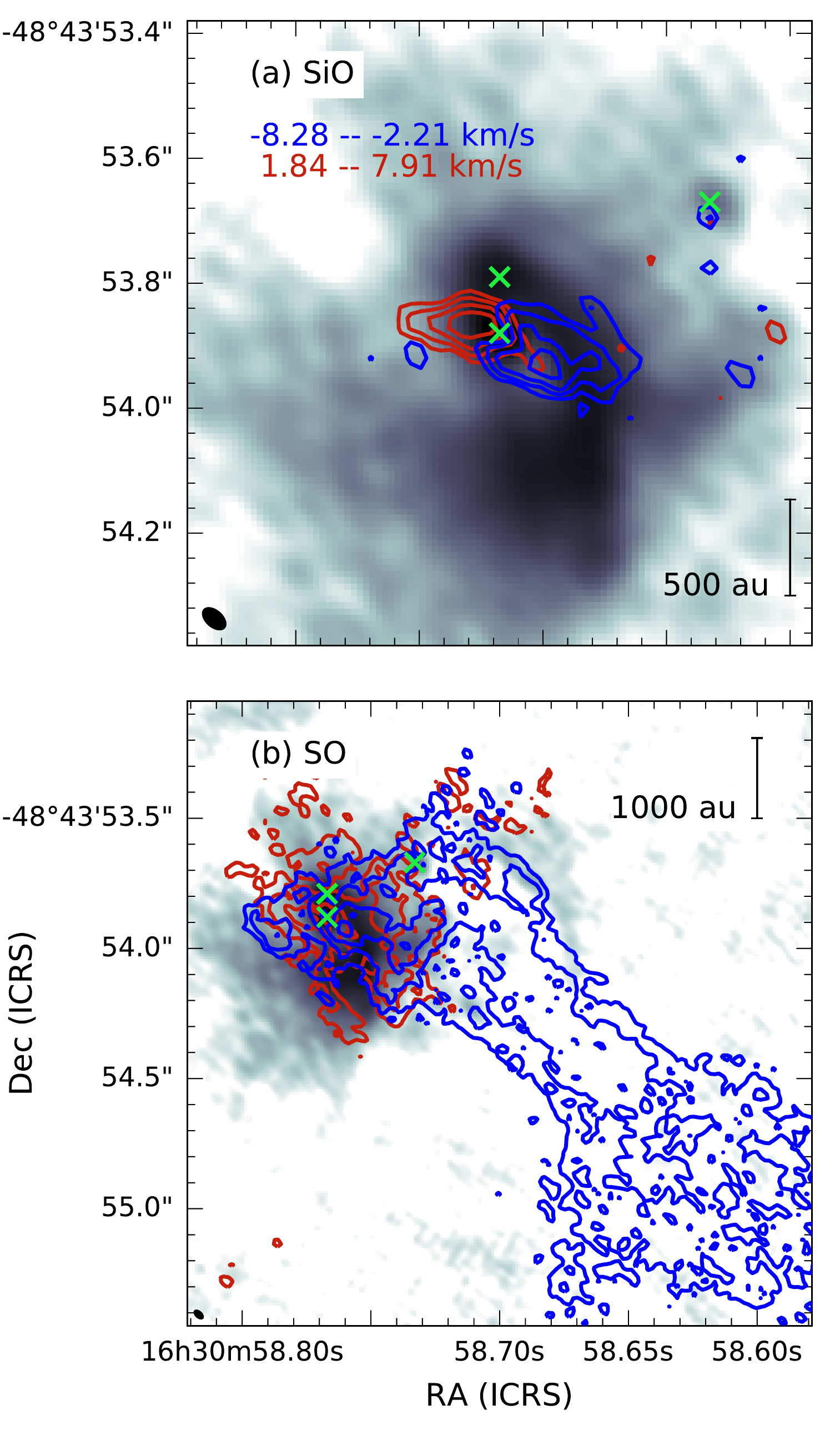}
\end{center}
\caption{Outflow emission from G335 ALMA1 in contours over the dust continuum.
(a) SiO $J=5-4$ red- and blue-shifted outflow lobes.
Contour levels are 5, 6, 7.2, $8.6\times \sigma_{\rm rms}$ with $\sigma_{\rm rms}=6$\,mJy\,beam$^{-1}$\,km\,s$^{-1}$.
(b) SO $J_K=6_5-5_4$ red- and blue-shifted emission.
Contour levels are 5, 7, 10, 14, $19\times \sigma_{\rm rms}$ with $\sigma_{\rm rms}=7.2$\,mJy\,beam$^{-1}$\,km\,s$^{-1}$.
The velocity integration range for each lobe is shown in (a).
The locations of the continuum sources are marked with green crosses.
The beam size is shown in the lower left corner.
}
\label{fig:sio:split}
\end{figure}

We use the SiO $J=5-4$ transition to study the outflow emission.
Figure~\ref{fig:sio:split} shows the maps from the red- and blue-shifted emission.
The zeroth order moment maps were calculated on the blue- and red-shifted sides of the channel closest to the line frequency.
We avoided the central 5 channels (2 channels at each side of the central one,  equivalent to $\sim1.2$\,km\,s$^{-1}$) and calculated the moments on a window of 10 channels ($\sim 6$\,km\,s$^{-1}$).
The origin of the outflow lobes is located around the source ALMA1a.
The direction of the blue-shifted emission is slightly inclined to the west (higher P.A.) in comparison to the large scale outflow shown in Figure~\ref{fig:sio:split}b from SO $J_K=6_5-5_4$ emission.
As expected for a disk-outflow system, the P.A. of the SiO blue-shifted lobe is close to perpendicular to the CH$_2$CHCN (disk) velocity gradient, and thus coincides with the rotation axis of the disk  (${\rm P.A.}=240\degr$; see Figure~\ref{fig:vinyl}d), while the large scale outflow P.A. is closer to 210\degr.
The emission producing the red-shifted lobe seems to come from gas escaping through a less dense region of the envelope as shown by the dip in dust continuum emission.

\begin{figure*}
\begin{center}
\includegraphics[scale=0.48]{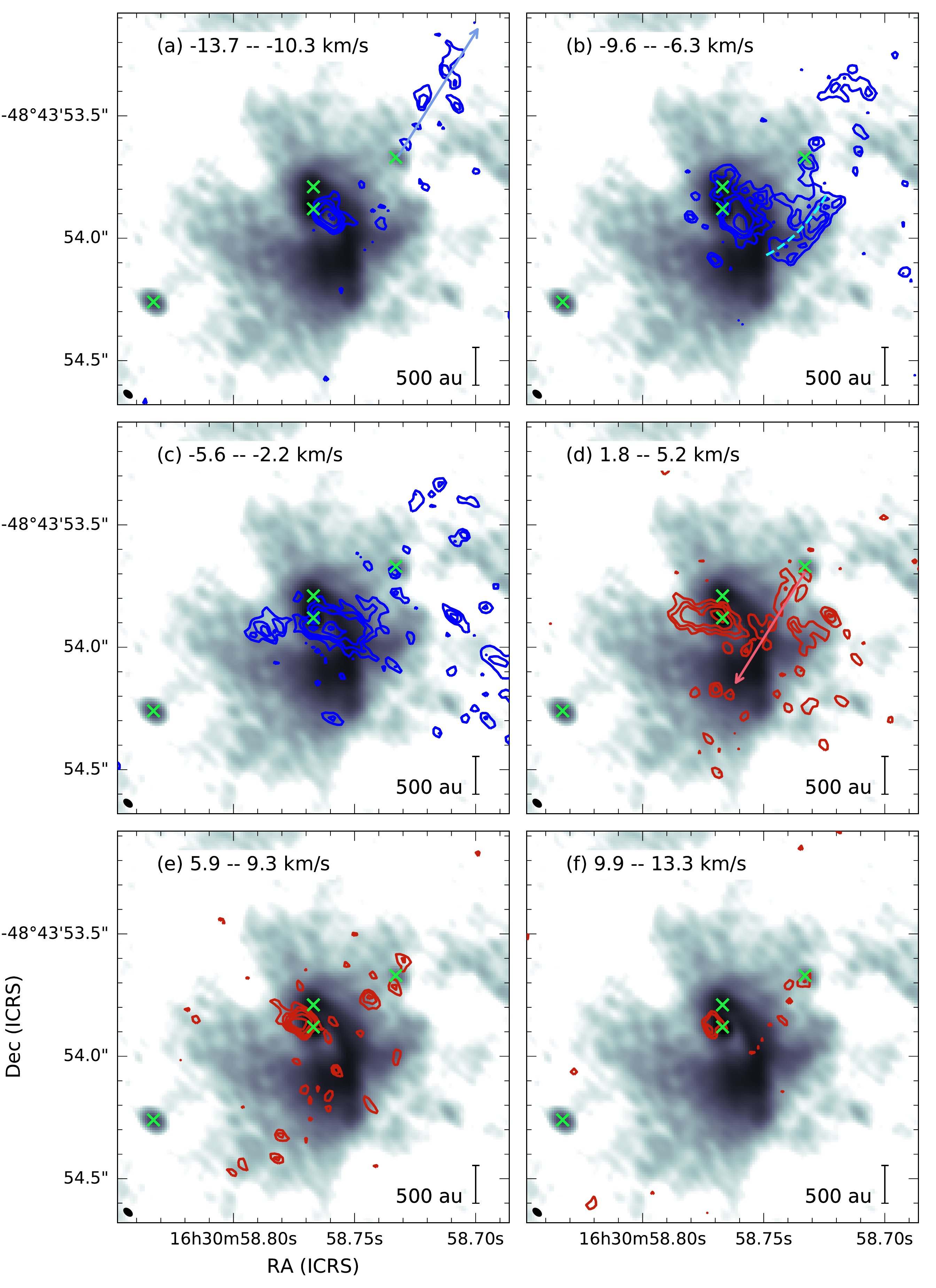}
\end{center}
\caption{Integrated intensities of SiO over different velocity ranges each side of the line center (contours) over dust continuum (gray scale).
(a)-(c) correspond to the blue-shifted emission and (d)-(f) to the red-shifted emission.
Contour levels are 3, 4, 5, 6.5, $8\times \sigma_{\rm rms}$ with $\sigma_{\rm rms}=4.5$\,mJy\,beam$^{-1}$\,km\,s$^{-1}$.
The blue arrow in (a) and the red arrow in (d) show the direction of the corresponding outflow lobes associated with ALMA1d.
The dashed cyan line in (b) indicates the location of a dust lane structure.
The velocity range of the moment maps is shown at the top of each frame, with zero systemic velocity equal to $v_{\rm LSR}=-46.9$\,km\,s$^{-1}$.
The locations of the continuum sources are marked with green crosses.
The beam size is shown in the lower left corner.
}
\label{fig:sio:rolling}
\end{figure*}

In order to study the interaction of the outflow with the circumstellar gas, we additionally calculated zeroth moment maps of groups of channels on each side of the line, i.e. similar to channel maps of averaged channels.
Each map consists of the zeroth moment maps of 5 channels ($\sim 3.3$\,km\,s$^{-1}$).
Figure~\ref{fig:sio:rolling} shows the maps on each side of the line and the details of the velocity range of each map.
High-velocity blue-shifted emission seems to trace the interaction of the outflow/jet with the envelope cavity.
In Figure~\ref{fig:sio:rolling}a and b, the emission is surrounded by the bow structure and is consistent with the location of wide CH$_3$OH lines towards the south-west.
Figure~\ref{fig:sio:rolling}b in addition shows SiO emission at the location of a dust lane structure (dashed cyan line).
Similarly, the high-velocity red-shifted emission is coming from regions closer to the source.

Figure~\ref{fig:sio:rolling} also shows evidence of an additional outflow associated to ALMA1d, with the blue-shifted outflow pointing in the north-west direction.
Given that both the blue- and red-shifted outflow emission appear in Figure~\ref{fig:sio:rolling}b, the systemic velocity of ALMA1d seems to be lower than that of ALMA1a.
Note that the C$^{18}$O emission in Figure~\ref{fig:lines}e is single peaked at roughly the same velocity of the ALMA1 region.
However, C$^{18}$O is likely tracing gas that belongs to the colder larger cloud rather than to ALMA1d in particular.
We thus report a velocity range in Table~\ref{tab:props} for source ALMA1d based on the range of Figure~\ref{fig:sio:rolling}b.
This difference between the systemic velocities indicates that the system ALMA1a/b and ALMA1d likely belong to the same parsec-scale association but the ALMA1d system is far enough such that there is no sign of interaction between their outflows.

\subsubsection{Infall signatures}

\begin{figure}
\begin{center}
\includegraphics[scale=0.48]{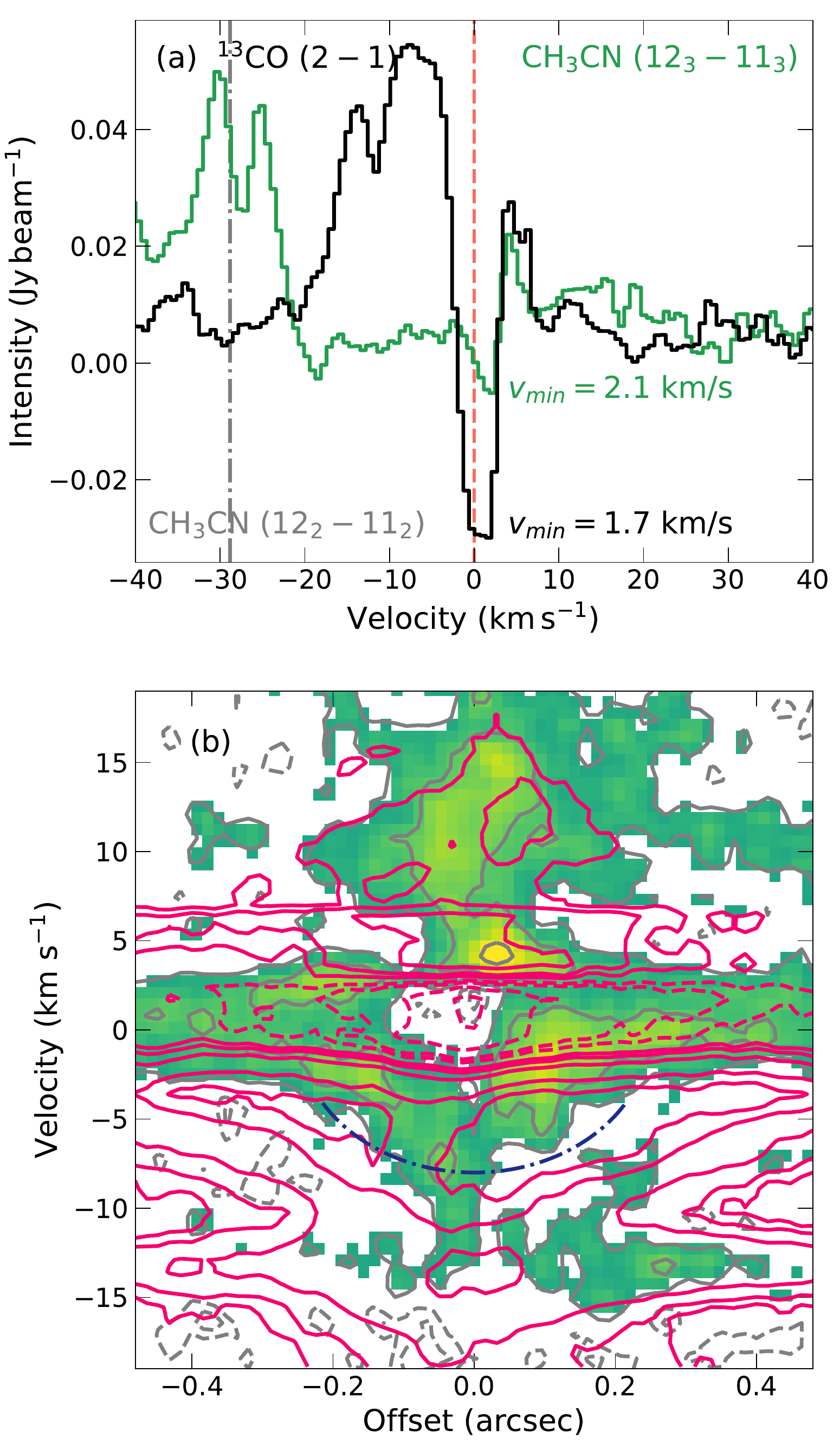}
\end{center}
\caption{Infall profiles and pv maps towards ALMA1a.
(a) CH$_3$CN $J_K=12_3-11_3$ (green line) and $^{13}$CO $J=2-1$ (black line) spectra towards ALMA1a.
The vertical dashed red line marks the zero velocity and the vertical dot-dashed grey line marks the position of the CH$_3$CN $J_K=12_2-11_2$ transition.
$v_{\rm min}$ corresponds to the velocity at the minimum of the absorption.
(b) pv maps of CH$_3$CN (colour and gray contours) and $^{13}$CO (red contours).
The CH$_3$CN contour levels are --4, --2, 4, 10, 20\,mJy\,beam$^{-1}$, while the $^{13}$CO contour levels are --30, --20, --10, --5, 5, 10, 20, 40, 50\,mJy\,beam$^{-1}$.
The color map only shows the data values over $3\sigma$ with $\sigma=1.2$\,mJy\,beam$^{-1}$.
The dot-dashed line show the ``C'' shape characteristic of infalling motions.
}
\label{fig:ch3cn}
\end{figure}

In addition to the blue central spot observed in CH$_3$OH, we observe that several lines that appear in emission in the compact configuration data become optically thick, resulting in inverse P-Cygni profiles.
In particular, the $K$ ladder of CH$_3$CN $J=12-11$ toward ALMA1a appear in absorption in the extended configuration data, while in the combined data the transitions with $K>2$ appear in absorption.
Figure~\ref{fig:ch3cn} shows the $K=3$ transition towards ALMA1a and the pv map from a slit of width 0\farcs05 (P.A.=150\degr) indicating the ``C'' shape characteristic of infall \citep{1997ApJ...488..241Z}.
The velocity shift of the line dip is slightly higher (0.4\,km\,s$^{-1}$) than that of $^{13}$CO $J=2-1$.
We fit a Gaussian function to the line absorption and estimate the error of the positions of the minima (see Appendix~\ref{ap:pcyg}).
We obtain a separation between the minimum value of the Gaussians of 0.8~\,km\,s$^{-1}$ and an error of 0.2~\,km\,s$^{-1}$.
The CH$_3$CN $J_K=12_3-11_3$ emission is likely tracing hotter gas than $^{13}$CO (see Table~\ref{tab:lines}), the shift indicates that the velocity of the infalling material is increasing towards the source \citep[e.g.,][]{2018A&A...615A.141B,2022arXiv220110438B}.
Additional observations with higher spectral resolution and/or better S/N are needed to separate these features further, thus allowing a detailed study of the infall motion.

\section{Discussion}
\label{sec:discussion}

\begin{figure*}
\begin{center}
\includegraphics[scale=0.48]{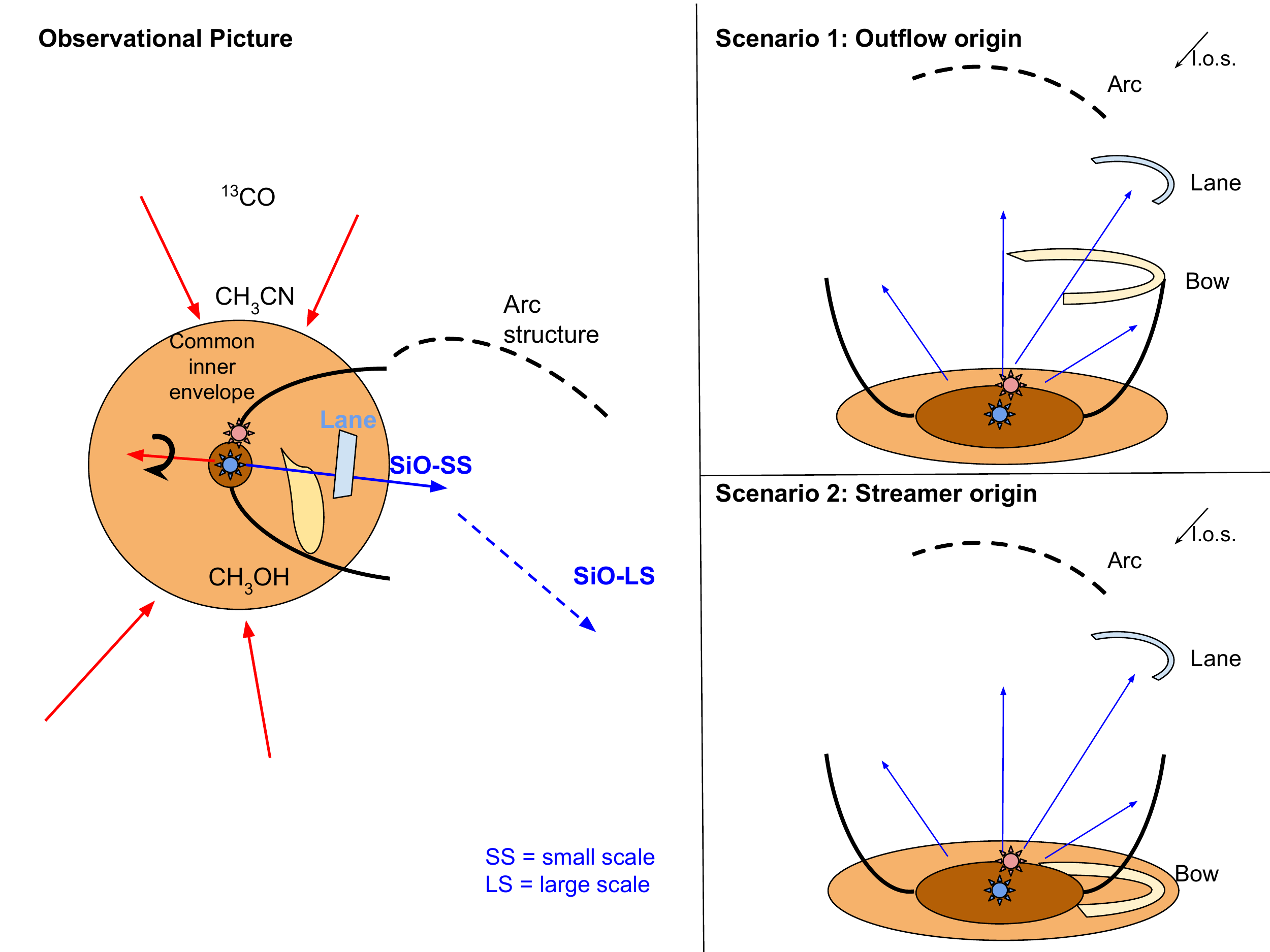}
\end{center}
\caption{Diagram summarizing the findings of Section~\ref{sec:results}, and the discussion in Section~\ref{sec:discussion} about the origin of the bow structure under an outflow origin and a streamer origin.
}
\label{fig:diagram}
\end{figure*}

Here we discuss the origin of the observed substructures, particularly the origin of the bow-shaped object.
Its origin will be framed into two paradigms:\\
1. A shocked region of the outflow cavity.\\
2. A streamer feeding the central region and/or forming a third stellar component.\\
The diagram in Figure~\ref{fig:diagram} summarizes the findings of the previous section and the two aforementioned paradigms.

\subsection{Outflow origin}

The location of the bow towards the south-west is consistent with large scale outflow direction (Figure~\ref{fig:sio:split}b).
Although the direction is not consistent with the bulk of the SiO emission (Figure~\ref{fig:sio:split}a), the structure may be the result of the interaction of a wide angle component of the outflow as indicated by the extended SiO emission in Figure~\ref{fig:sio:rolling}b.
The interaction of the outflow and the envelope material can produce an accumulation of gas resulting in the observed structure.
Such structures are predicted by simulations \citep[e.g.,][]{2015ApJ...800...86K,2018A&A...616A.101K} and have been observed in other high-mass star-forming regions \citep[e.g.,][]{2003A&A...412..735P}.

The presence of unresolved free-free radio emission towards this region \citep[${\sim}2\arcsec$ resolution][]{2015A&A...577A..30A} and the observations presented here, where the continuum emission from different components can be separated, begs the question of whether the radio emission is produced by a jet rather than an HC \ion{H}{2} region.
We revisit the fit to the sub-mm/radio SED of \citet[][]{2015A&A...577A..30A} to test whether the bow emission (red contour level in Figure~\ref{fig:continuum}) is produced by a jet.
Here we assume that the free-free emission comes from ionized gas in the bow, and fit a power law with a dust and a free-free component to the SED:
\begin{equation}
    S_\nu = S_d \left( \frac{\nu}{\nu_0} \right)^{\beta+2} + S_{ff} \left( \frac{\nu}{\nu_0} \right)^{\alpha}~,
\end{equation}
where $\beta$ is the dust emissivity index and $\alpha$ the free-free spectral index.
Since we lack data at similar angular resolution at higher frequencies to fit the dust emissivity index, we assume $\beta=1.97$ from \citet{2015A&A...577A..30A}.
Figure~\ref{fig:sed} shows the results of the fit.
The fitted free-free spectral index of 1.5 is relatively insensitive to dust emissivity indices in the 1.5--2.0 range.
This value is slightly smaller than that derived by \citet{2015A&A...577A..30A}, $\alpha=1.67-1.9$.
A free-free spectral index of 1.5 is not common in jets (but plausible) and particularly not in shock ionized jets, where indices $\alpha<0$ are expected \citep[e.g.,][]{2021MNRAS.504..338P}.

The extended tentative H30$\alpha$ emission is also not consistent with jet emission which is more compact and closer to the source \citep[e.g.,][]{Guzman2020}.
The H30$\alpha$ emission is more consistent with \ion{H}{2} regions, but the extension of the emission would indicate that the source is more evolved than previously estimated by \citet{2015A&A...577A..30A,2021A&A...645A.142A}.

\subsection{Streamer origin}

Streamers feeding protostellar high-mass systems have been observed in a few other regions at 1000 - 10000 au scales \citep[e.g.,][]{2017MNRAS.467L.120M,Izquierdo18,2020ApJ...905...25G,2021ApJ...915L..10S}. 
The observations presented here show this type of structures at smaller scales ($<1000$\,au), which have been observed in disks of single systems \citep[e.g.,][]{2019A&A...627L...6M,2020A&A...634L..11J}.
Simulations show that gravitational instabilities of the circumstellar disk can produce spiral-like arms that can feed the protostar in episodic bursts of accretion \citep[e.g.,][]{2018MNRAS.473.3615M,2021A&A...652A..69M,2021MNRAS.507.6061R}.
These simulations show over-densities with shapes similar to that of the continuum observations presented here, and are predicted to be observable by ALMA at high-resolution \citep{2019A&A...632A..50A,2019MNRAS.487.4473M}.

The stability of a spiral structure can be determined from the Hill criterion, which relates the self-gravity of the spiral structure with the shear forces exerted over it \citep{2012MNRAS.423.1896R,2018MNRAS.473.3615M}.
An unstable spiral would fragment and consequently form another star/planet, while a stable filament would continue to funnel gas to the disk interior \citep{2012MNRAS.423.1896R}.
A spiral arm is unstable if the spiral width $l<2R_{\rm Hill}$ \citep{2012MNRAS.423.1896R}, with the Hill radius given by \citep{2012MNRAS.423.1896R}
\begin{equation}
    R_{\rm Hill} = \left( \frac{G\Sigma l^2}{3\Omega^2} \right)^{1/3}\, ,
\end{equation}
with $\Sigma$ the surface density of the spiral, $l$ the spiral width, and $\Omega$ the rotation rate.
The surface density derived from the dust emission is $\Sigma= M_d/(\pi r^2)=42$\,g\,cm$^{-2}$, while the bow width is $l=0\farcs11=360$\,au.
This width is measured across the 130$\sigma$ level and passing through the peak emission in the radial direction with respect to ALMA1a.
Width values range between ${\sim}0\farcs02$ (65\,au) up to 0\farcs14 (455\,au).
From the rotation velocity $v_\phi$, we obtain a rotation rate of
$\Omega=|v_{\phi,obs}|/(r\sin i)=1.0\times10^{-10}$\,s$^{-1}$ for an inclination angle of 26\degr\ (see Section~\ref{sec:results:kin}).
This rotation rate is more than twice higher than those obtained in the line modeling presented in \citetalias{2021ApJ...909..199O} ($\lesssim 4\times10^{-11}$\,s$^{-1}$).
With these values we obtain a Hill radius of $R_{\rm Hill}=92$\,au.
This Hill radius implies that the spiral is stable.
Governed by the gravity of the central sources, the spiral would be destined to be accreted by the binary system.

Figure~\ref{fig:methanol:cassis}c shows a velocity gradient along the spine of the bow structure with values ranging from ${\sim}0.8$\,km\,s$^{-1}$ near the peak continuum emission to ${\sim}-0.3$\,km\,s$^{-1}$ closer to the binary sources.
This velocity gradient, however, may be produced by the large scale common disk rotation and/or contaminated by the outflow motion observed in the same direction.
The lack of a velocity gradient in the bow structure from other molecular tracers due to weak line emission is the main caveat to confirm the streamer scenario.
This is likely caused by the dust becoming optically thick at smaller scales (see Section~\ref{sec:results:props}).
However, the presence of, e.g., C$^{18}$O absorption in Figure~\ref{fig:lines}f indicates that there is a temperature gradient which is not considered in the Hill criterion.
This temperature gradient may be caused by an already formed additional source, or heating from the working surfaces of an accretion shock.
On the other hand, under the outflow origin, the gradient can be produced by heating from the working surfaces of an outflow shock.
In the cases of accretion and outflow shocks, accumulation of gas would make the dust optically thick.

\section{Conclusions}
\label{sec:conclusions}

We analyzed high-resolution ALMA 1.3\,mm observations of the high-mass source G335.579--0.272 MM1 ALMA1 that resolve scales of ${\sim}200$\,au.
The continuum observations reveal 4 sources inside the region, with ALMA1a and ALMA1b likely forming a binary system.
We detect a fifth continuum peak located to the south-west of the binary system with a bow shape and connected to the main sources by continuum emission.
These three sources are located at the center of the gravitational potential well of the ALMA1 region.

Line emission is damped towards the binary sources and the bow, and appears in absorption in many common hot core lines (e.g., CH$_3$CN).
These indicate that lines are becoming optically thick, and given the bright continuum some are presenting inverse p-Cygni profiles.
Emission from CH$_3$OH transitions traces a common disk/inner envelope of ${\sim}3000$\,au diameter. 
Part of the emission is likely coming from gas in the blue-shifted outflow cavity, where lines are wider.
The surrounding emission shows a blue-shifted central region surrounded by  red-shifted emission, characteristic of infalling matter when lines are becoming optically thick.

The lack of line emission tracing the kinematics of ALMA1b precludes the determination of the parameters from the binary system. 
Nonetheless, the CH$_2$CHCN emission tracing the gas around ALMA1a allows us to determine its rotation direction and estimate a Keplerian mass of 3\,\msun\ under an estimated inclination angle of the system of 26\degr.
These inclination estimate is based on geometrical considerations about the shape of the CH$_2$CHCN emission, and is smaller than previous estimates from outflows which indicates precession of the sources.

ALMA1a is powering the outflow as shown by SiO emission.
The direction of the outflow close to the source is consistent with the rotation axis as derived from the CH$_2$CHCN velocity gradient (P.A.=240\degr).
SiO features are consistent with features in the continuum, indicating the location of previous interactions between the outflow and the envelope cavity walls.

We explore the origin of the bow.
While its location and outflow line emission supports an outflow origin for the structure as matter swept and/or shocked along the the outflow cavity walls, there is not enough evidence to discard an infalling streaming origin.
The latter could be accreted by the central source(s) in burst or even form a third companion.
Additional multi-wavelength observations at similar resolution are necessary to assess the bow origin. 

\acknowledgments
The authors would like to thank the anonymous referee for the thoughtful feedback.
F.O. and H.-R.V.C. acknowledge the support of the Ministry of Science and Technology of Taiwan, projects No. 109-2112-M-007 -008 -, 110-2112-M-007 -023 - and 110-2112-M-007 -034 -.
P.S. was partially supported by a Grant-in-Aid for Scientific Research (KAKENHI Number 18H01259) of the Japan Society for the Promotion of Science (JSPS). 
A.G. acknowledges support from NSF grant AST 2008101.
This paper makes use of the following ALMA data: ADS/JAO.ALMA\#2016.1.01036.S. ALMA is a partnership of ESO (representing its member states), NSF (USA) and NINS (Japan), together with NRC (Canada), MOST and ASIAA (Taiwan), and KASI (Republic of Korea), in cooperation with the Republic of Chile. The Joint ALMA Observatory is operated by ESO, AUI/NRAO and NAOJ. \\

\facility{ALMA}

\software{CASA \citep{2007ASPC..376..127M}, astropy \citep{astropy:2013,astropy:2018}, GoContinuum \citep{2020zndo...4302846O}, YCLEAN \citep{2018zndo...1216881C} }

\vspace{50mm}

\newpage
\vspace{80mm}
\appendix

\section{Additional Figures}

\begin{figure}[h]
\begin{center}
\includegraphics[scale=0.6]{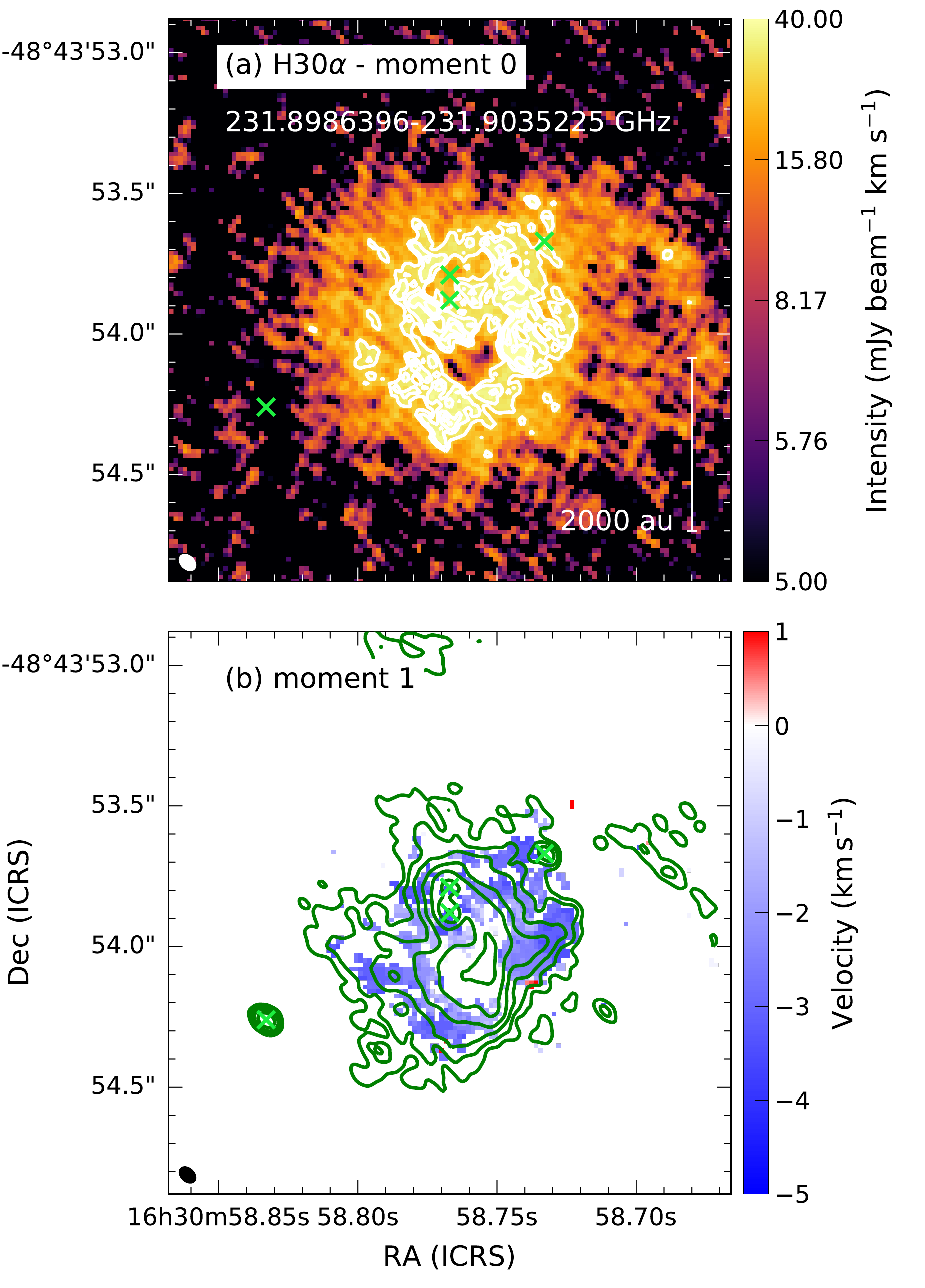}
\end{center}
\caption{Tentative H30$\alpha$ emission in G335 ALMA1.
(a) Zeroth order moment map.
Contour levels are 5, 6, 7, $8\times\sigma_{\rm rms}$ with $\sigma_{\rm rms}=5.4$\,mJy\,beam$^{-1}$\,km\,s$^{-1}$.
The frequency range shown in the figure corresponds to the frequency integration range in the rest frame.
(b) First moment map in color scale.
Zero systemic velocity corresponds to the source $v_{\rm LSR}$ (--46.9\,km\,s$^{-1}$).
Contours correspond to the continuum emission (see Figure~\ref{fig:continuum}b).
The beam is shown in the lower left corner.
}
\label{fig:ap:halpha}
\end{figure}

\begin{figure}[h]
\begin{center}
\includegraphics[scale=0.48]{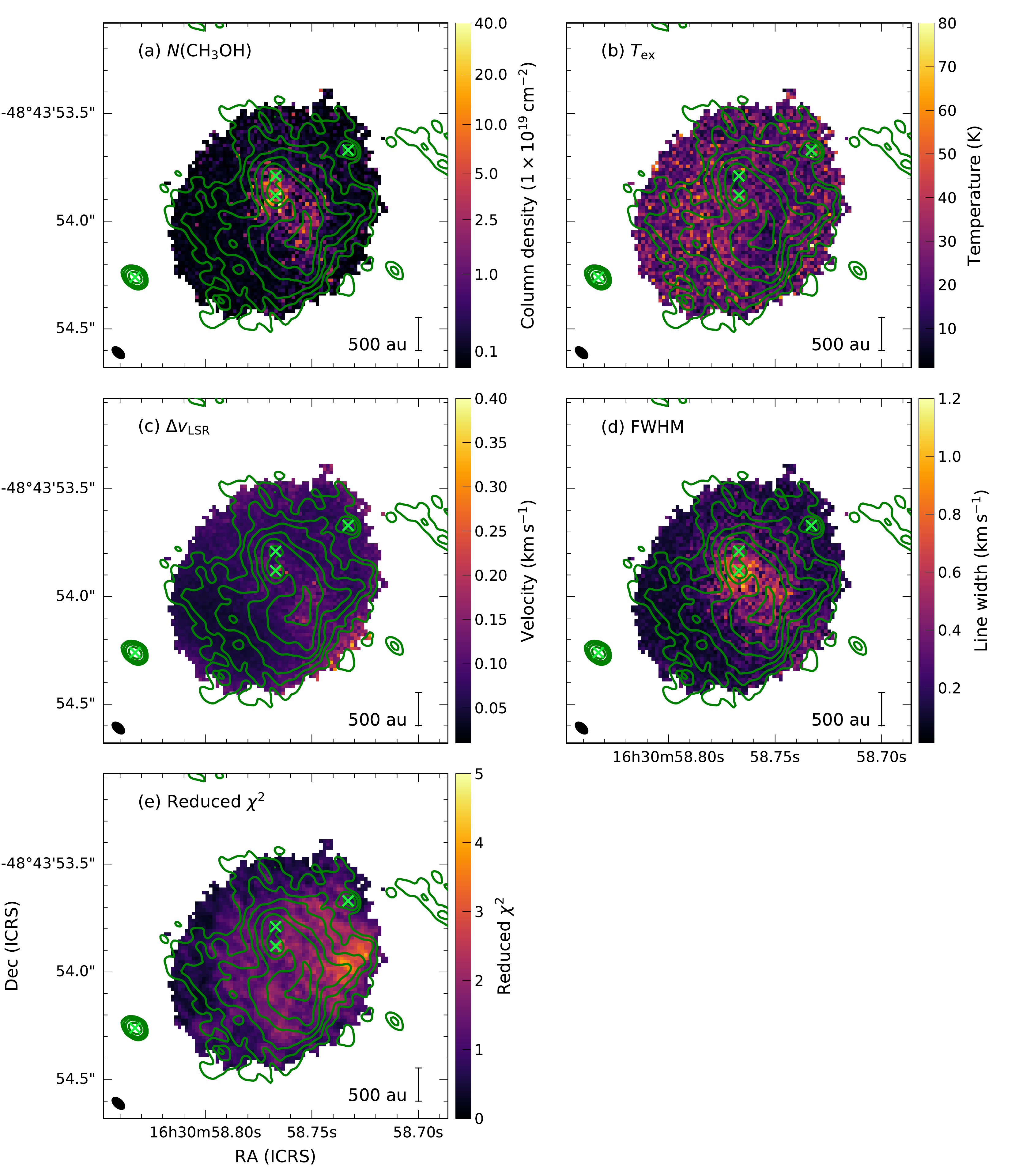}
\end{center}
\caption{CH$_3$OH LTE modeling standard deviation maps and reduced $\chi^2$ map.
(a) CH$_3$OH column density error map.
(b) Excitation temperature error map.
(c) Velocity shift with respect to the systemic velocity error map.
(d) Line FWHM error map.
(e) $\chi^2$ map
Contour levels correspond to 5, 8, 12, 16, $20\times \sigma_{\rm rms}$ with $\sigma_{\rm rms}=11$\,mJy\,beam$^{-1}$.
The green contours correspond to the continuum emission from the extended configuration, Figure~\ref{fig:continuum}(b). 
The location of the continuum sources are marked with green crosses.
The beam size is shown in the lower left corner. 
}
\label{fig:ap:methanol:cassis:error}
\end{figure}

\begin{figure}[h]
\begin{center}
\includegraphics[scale=0.7]{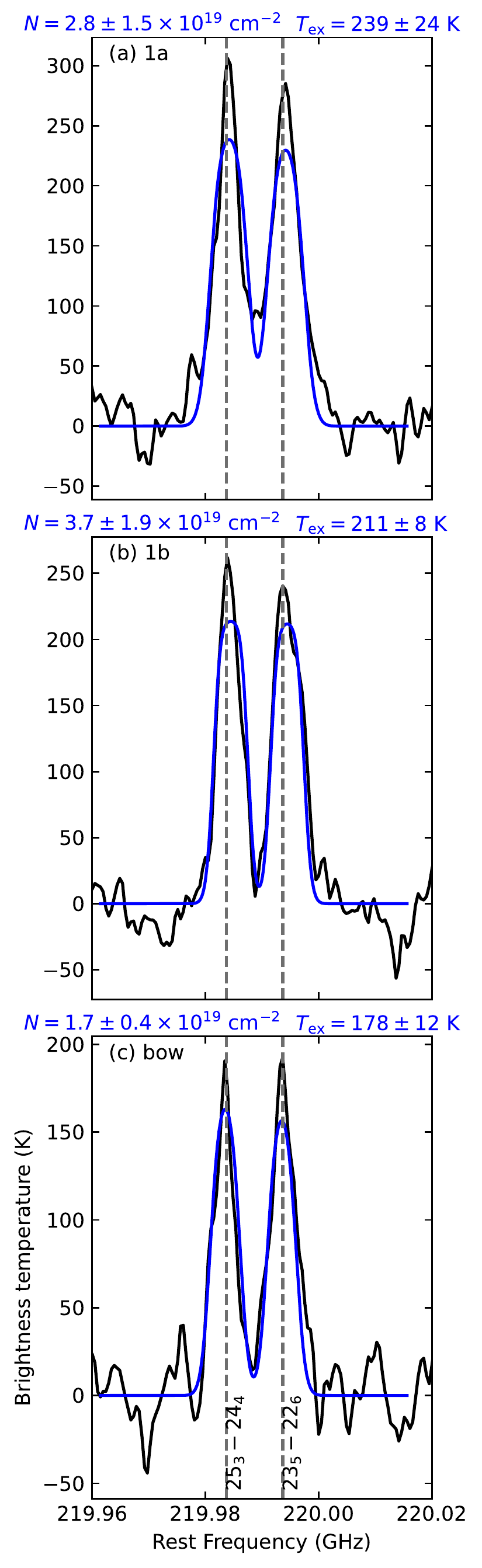}
\end{center}
\caption{CH$_3$OH line emission towards the continuum peak of (a) ALMA1a, (b) ALMA1b, and (c) bow structure.
Observations are presented in black lines, while the best-fitting model is shown in blue lines.
The column density, $N$, and excitation temperature, $T_{ex}$, of the best-fitting model and $1\sigma$ errors are shown over each panel.
Vertical dashed lines correspond to the transitions frequency.
}
\label{fig:ap:methanol}
\end{figure}

\begin{figure}[h]
\begin{center}
\includegraphics[scale=0.7]{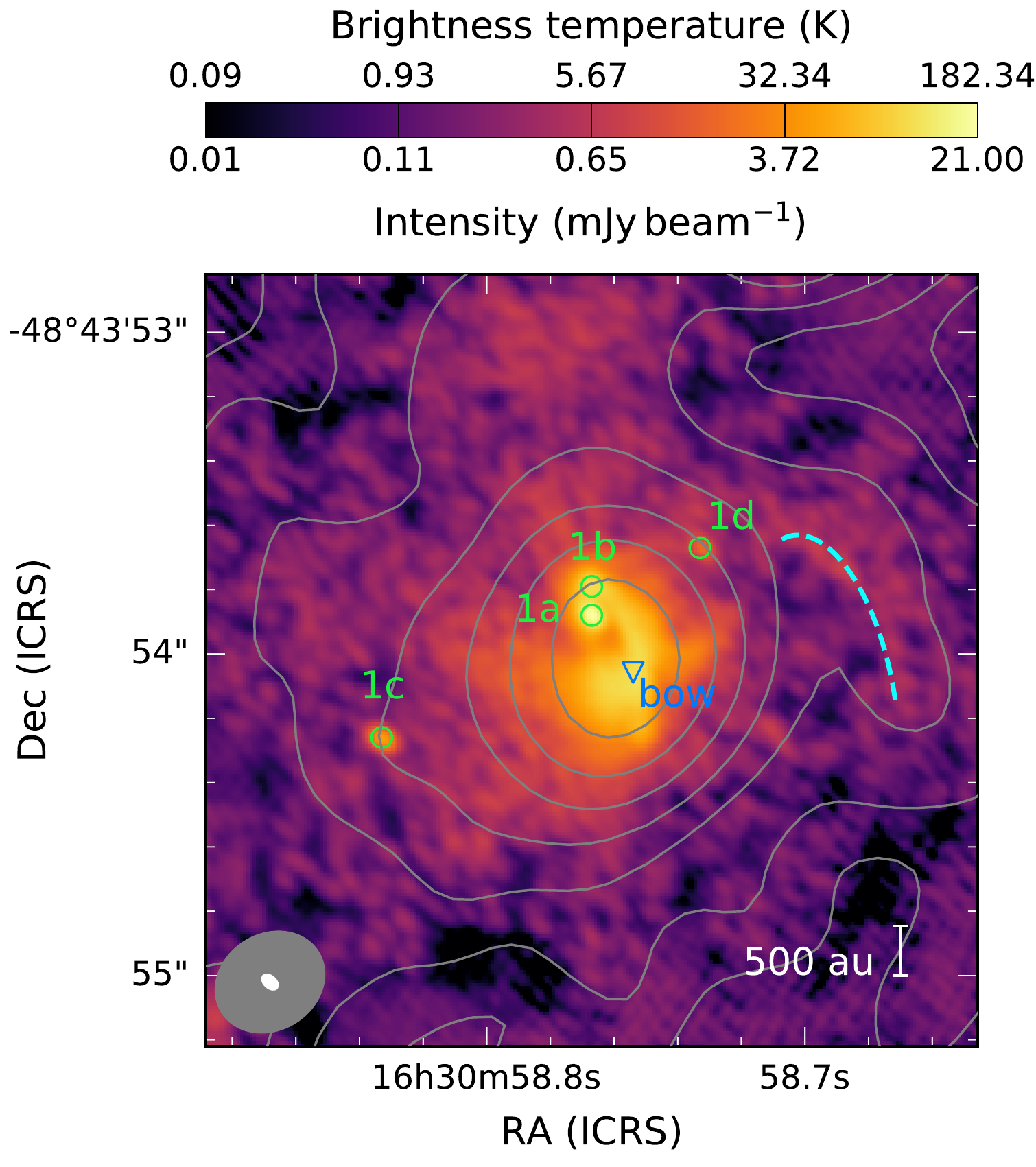}
\end{center}
\caption{Continuum maps of ALMA1 from the combined configuration (color scale) and compact configuration from \citetalias[][]{2021ApJ...909..199O} (gray contours).
Contour levels are 5, 10, 20, 40, 80, 160, and 320$\times\sigma_{\rm cont}$
with $\sigma_{\rm cont} =0.4$\,mJy\,beam$^{-1}$ and a beam size of $0\farcs36\times0\farcs30$.
Continuum sources are marked with green circles.
The dashed line shows the arc-like structure detected in the compact configuration data.
The beam of the compact configuration is shown in the lower left corner in gray while the beam of the combined configuration is shown in white.
}
\label{fig:ap:alma1}
\end{figure}

\begin{figure}[h]
\begin{center}
\includegraphics[scale=0.7]{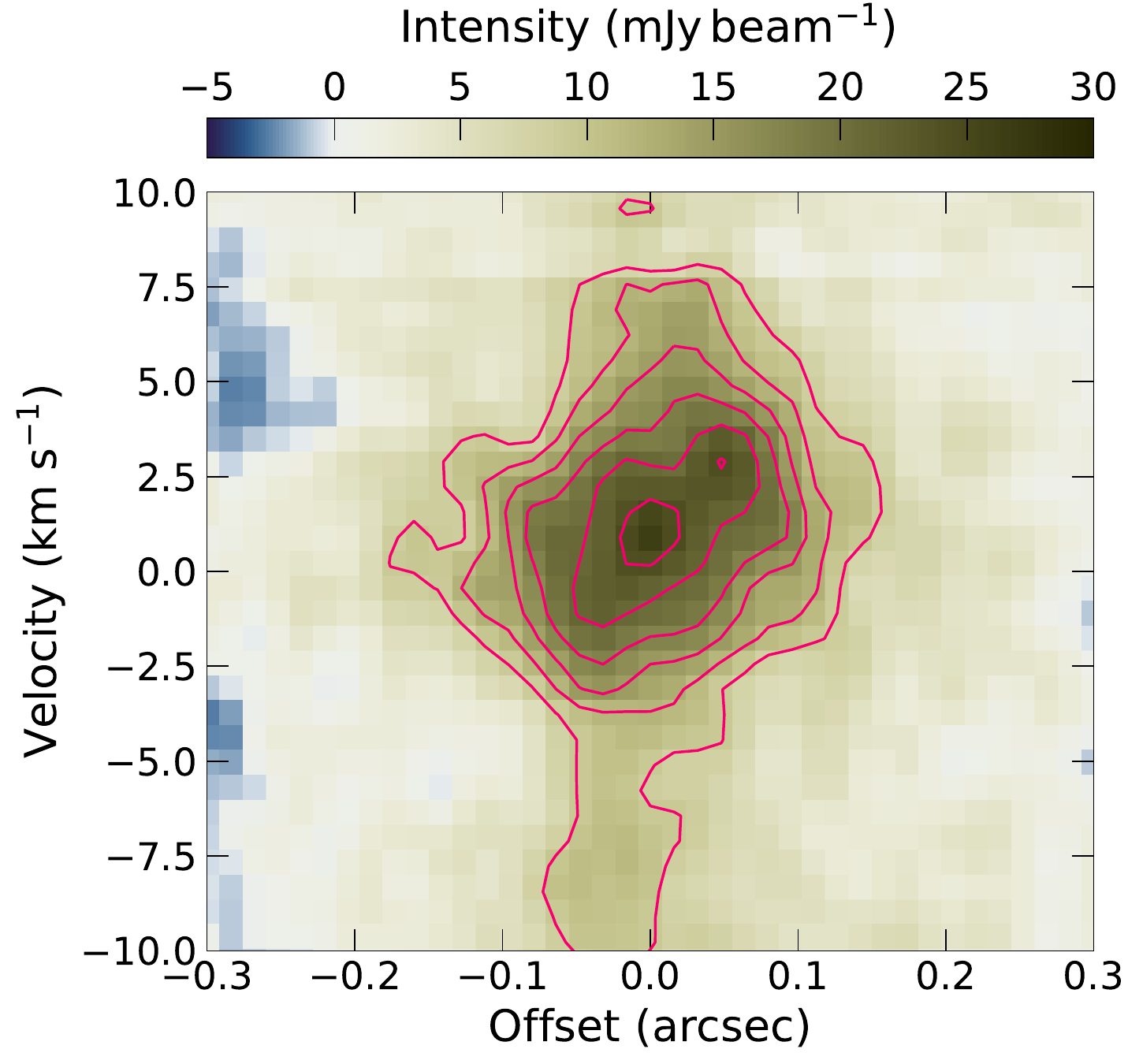}
\end{center}
\caption{Position velocity map of CH$_2$CHCN emission from the combined data set.
Contour levels are 3, 4, ... $8\times\sigma$ with $\sigma=3$\,mJy\,beam$^{-1}$.
The map was calculated in a slit centered in the position of ALMA1a with a ${\rm P.A.}=150\degr$ and a width of 0\farcs05.}
\label{fig:ap:pvmap}
\end{figure}

\section{Inverse P-Cygni Gaussian Fit}
\label{ap:pcyg}

\begin{figure}[h]
\begin{center}
\includegraphics[scale=0.7]{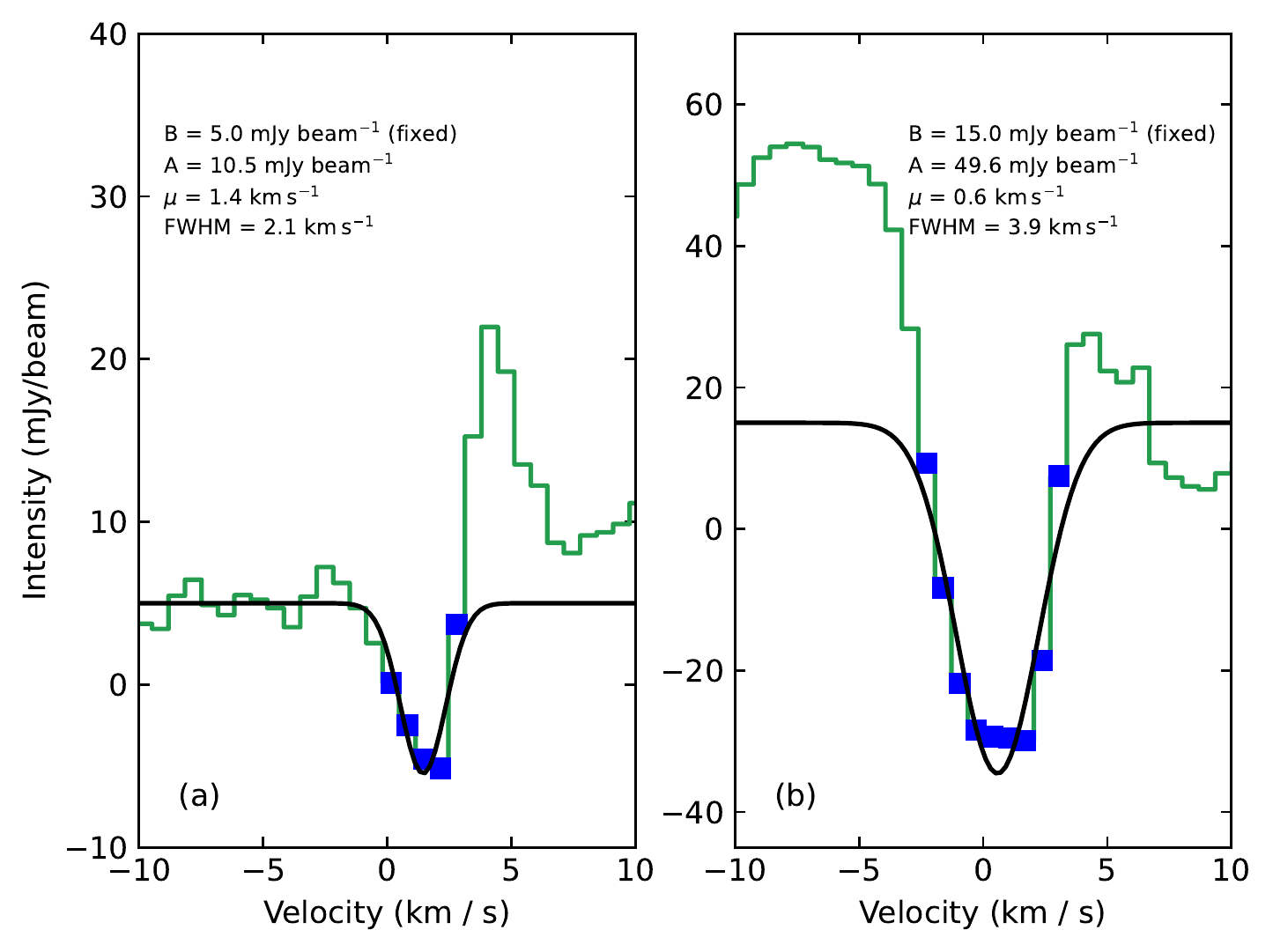}
\end{center}
\caption{Gaussian fit to the inverse P-Cygni profile of (a) CH$_3$CN $J=12-11$ $K=3$ and (b) $^{13}$CO $J=2-1$ towards G335 ALMA1a.
The green line shows the data in Figure~\ref{fig:ch3cn}, blue squares correspond to the fitted points and black line shows the fit.}
\label{fig:ap:pcygni:fit}
\end{figure}

In order to estimate the statistical significance of the distance between the minima of the inverse P-Cygni of CH$_3$CN and $^{13}$CO, we first fit a Gaussian to the intensity of the form:
\begin{equation}
    I(v) = B - A \exp{\left(-\frac{(v - \mu)^2}{2\sigma^{2}}\right)} .
\end{equation}
We limit the fit to the points in the absorption profile and fix the baseline, $B$.
Figure~\ref{fig:ap:pcygni:fit} shows these fitted data and the values of the best fit.

To estimate the error in the value of the velocity at the minimum, $\mu$, we adapt the expected position uncertainty equation for astrometric measurements in \citet{1988ApJ...330..809R}, as:
\begin{equation}
    \sigma_\mu = \left(\frac{4}{\pi}\right)^{0.25} \frac{\sigma_{\rm rms}}{A} \sigma
\end{equation}
with $\sigma_{\rm rms}=2$\,mJy\,beam$^{-1}$ the noise per channel.
We obtain a $\sigma_\mu = 0.2$\,km\,s$^{-1}$ and $0.1$\,km\,s$^{-1}$ for CH$_3$CN and $^{13}$CO, respectively.

\bibliography{manuscript}{}
\bibliographystyle{aasjournal}

\end{document}